\documentclass[12pt]{article}
\pdfoutput=1

\usepackage{amssymb,amsmath}
\usepackage{mathdots}

\usepackage{cite}
\usepackage[pdftex]{graphicx}

\usepackage{bbm}
\usepackage{multirow}
\usepackage{booktabs}

\newcommand{\otoprule}{\midrule[\heavyrulewidth]}

\def\ft{\tfrac}

\def\ov{\overline}
\def\ie{{\it i.e.\ }}

\newcommand{\be}{\begin{equation}}
\newcommand{\ee}{\end{equation}}
\newcommand{\bea}{\begin{eqnarray}}
\newcommand{\eea}{\end{eqnarray}}
\newcommand{\ba}{\begin{array}}
\newcommand{\ea}{\end{array}}
\def\nn{\nonumber}

\newcommand{\drawsquare}[2]{\hbox{%
\rule{#2pt}{#1pt}\hskip-#2pt
\rule{#1pt}{#2pt}\hskip-#1pt
\rule[#1pt]{#1pt}{#2pt}}\rule[#1pt]{#2pt}{#2pt}\hskip-#2pt
\rule{#2pt}{#1pt}}

\newcommand{\fund}{\raisebox{-.5pt}{\drawsquare{6.5}{0.4}}}

\newcommand{\Z}{{\mathbb Z}}

\newcommand{\C}{{\mathbb C}}
\def\Cint{{\mathbb C}}
\def\Zint{{\mathbb Z}}

\begin{document}
\begingroup
\font\cmss=cmss10 \font\cmsss=cmss10 at 7pt

\vskip -0.5cm
\rightline{\small{\tt ROM2F/2013/10}}
\rightline{\small{\tt DFPD-13/TH/12}}

\vskip .7 cm
\endgroup
\hfill
\vspace{18pt}
\begin{center}
{\Large \textbf{Unoriented Quivers with Flavour}}
\end{center}

\vspace{6pt}
\begin{center}

  {\textsl{ Massimo Bianchi $^{\dagger}$ \footnote{\scriptsize \tt massimo.bianchi@roma2.infn.it}, Gianluca Inverso $^{\dagger}$ \footnote{\scriptsize \tt gianluca.inverso@roma2.infn.it}, Jose Francisco Morales $^{\dagger }$ \footnote{\scriptsize \tt francisco.morales@roma2.infn.it}  \\ \& Daniel Ricci Pacifici $^\ddagger$ \footnote{\scriptsize \tt daniel.riccipacifici@pd.infn.it}}}
  
\vspace{1cm}
$\dagger$ \textit{\small I.N.F.N. Sezione di Roma ``TorVergata'' \&\\  Dipartimento di Fisica, Universit\`a di Roma ``TorVergata", \\
Via della Ricerca Scientifica, 00133 Roma, Italy }\\  \vspace{6pt}
$ \ddagger $ \textit{\small Dipartimento di Fisica e Astronomia, Universit\`a degli Studi di Padova \&\\  I.N.F.N, Sezione di Padova,\\
Via Marzolo 8, 35131, Padova, Italy}\\  \vspace{6pt}

\end{center}

\begin{center}
\textbf{Abstract}
\end{center}
We discuss unoriented quivers with flavour that arise from D3-branes at local orbifold singularities, in the presence of $\Omega$-planes and non-compact D7-branes. We produce a wide class of unoriented quiver gauge theories, including new instances of ${\cal N}=1$ superconformal theories.
We then consider unoriented D-brane instanton corrections of both  `gauge' and `exotic' kinds.
In particular, we show that conformal symmetry can be dynamically broken via the generation of exotic superpotentials. 
Finally we discuss aspects of the recently proposed ${\cal N}=1$  remnant of ${\cal N}=4$ S-duality.
We identify new candidate dual pairs for the $\C^3/\Z_n$  series of unoriented quiver gauge theories with $n$ odd.
 
\vspace{4pt} {\small

\noindent }

\newpage

\tableofcontents

\section{Introduction}

A large class of ${\cal N}=1$ superconformal field theories arise from D3-branes transverse to Calabi--Yau singularities. The near-horizon geometry is AdS$_5\times X$ where $X$ is a Sasaki--Einstein space, that is the base of a non-compact CY cone \cite{Klebanov:1998hh,Morrison:1998cs,Benvenuti:2004dy,Franco:2005sm}. 
Particular attention has been devoted to local orbifold and more general toric singularities, since the resulting quiver theories admit an elegant  description in terms of brane tilings and dimers, that encode their low-energy dynamics and their moduli spaces \cite{Hanany:2005ve,Franco:2005rj, Hanany:2011iw, Hanany:2012hi, Hanany:2012vc}.
Less is known about the inclusion of orientifold planes and flavour branes, since both typically break superconformal invariance (see \cite{Franco:2007ii} for previous work on unoriented brane tilings and dimers). 

On the other hand, configurations with orientifold planes and flavour branes  provide us with concrete examples of semi-realistic models for particle physics \cite{Angelantonj:1996uy, Aldazabal:2000sa,Cvetic:2001tj} (see \cite{Blumenhagen:2006ci} for a review and references therein). In the case where the brane system is located at the fixed point of an orientifold involution, the low energy  dynamics is governed by a local unoriented quiver theory whose quantum consistency relies on local tadpole cancellation and  admits a full-fledged world-sheet description.

Here we mainly focus on the case of $\C^3/\Z_n$ singularities with fractional D3-branes, non-compact  D7-branes\footnote{Brane tilings with flavour have been recently considered in  \cite{Forcella:2008au,Franco:2012mm, Franco:2012wv}.} and $\Omega$-planes of general type.
We will rederive the various consistency conditions, most notably the relation between twisted tadpoles and anomalies in presence of flavour branes \cite{Aldazabal:1999nu, Bianchi:2000de,Uranga:2000xp}.  The gauge group will be a product of unitary, orthogonal and symplectic groups. Matter will appear in fundamental, symmetric or anti-symmetric representations. In particular, we show that the presence of flavour branes allows for a rich pattern of quiver theories including new instances of ${\cal N}=1$ superconformal theories.  We will also discuss D-brane instanton corrections of both kinds, `gauge' and `exotic', related to instantons sitting in an occupied or an empty node of the quiver, respectively \cite{Billo:2002hm,Bianchi:2007wy, Argurio:2007vqa, Bianchi:2009bg, Blumenhagen:2009qh, Bianchi:2009ij,Bianchi:2012ud}.
Interestingly we find superconformal theories whereby instanton induced superpotentials break conformal symmetry in a dynamical fashion.  
Finally we discuss aspects of the new ${\cal N}=1$ strong-weak coupling duality, proposed by \cite{GarciaEtxebarria:2012qx} as a remnant of  ${\cal N}=4$ S-duality.
In particular we will identify new candidate dual pairs and propose that the duality relation can be understood in purely geometric terms.

The plan of the paper is as follows. In section \ref{sunoriented} we describe the spectrum of the quiver theories and present general formulas for the one-loop anomalies and tadpoles entirely written in terms of the intersection numbers  codifying the singularity (quiver diagram).
In section \ref{sconformal} we show that the presence of flavour branes allows for new instances of ${\cal N}=1$ superconformal quiver gauge theories.
Besides a number of truly superconformal quiver theories we find an infinite class of theories where breaking of conformal symmetry shows up only in the running of the coupling associated to an empty node. 
In section \ref{sinstanton} we study the effects of D-brane instantons of both kinds: `gauge' and `exotic'. 
We show in particular that conformal symmetry can be broken in a dynamic fashion via the generation of exotic superpotentials. Finally in
section \ref{sdual} we propose an infinite series of new candidates for ${\cal N}=1$ strong-weak pairs of dual quiver gauge theories. We collect in Appendix~A a self-contained discussion of the Klein-bottle, Annulus and Moebius-strip one-loop amplitudes, anomalies and tadpoles of $\C^3/\Z_n$ orientifold theories. 

\bigskip
\noindent{\it Note added}

\noindent While this paper was being typewritten a related interesting paper by S.~Franco and A.~Uranga \cite{Franco:2013ana}
appeared that discusses flavour D7-branes in general bipartite field theories, yet without the inclusion of $\Omega$-planes.

\section{IIB on $\C^3/\Z_n$ orientifolds }
\label{sunoriented}

We are interested in unoriented quiver theories living on D3-branes at  $\C^3/\Gamma$ singularities with $\Gamma$ a discrete and abelian group. We start by considering the case $\Gamma=\Z_n$, the main focus of our analysis.  We denote by $X^I$, $I=1,2,3$, the complex coordinates of $\C^3$ and by $\Theta $ the generator of the $ \Z_{n}$ orbifold group action
\be
\Theta: \quad X^I \to w^{a_I} X^I, \qquad w=e^{\frac{2\pi  {\rm i}}{n}},
\ee
with $a_I$ integers satisfying the supersymmetry-preserving Calabi--Yau condition 
\be \sum_{I=1}^3  a_I=0  \quad ({\rm mod} \:  n). \ee

The orbifold action has a single fixed point at the origin. Before the inclusion of D-branes and the $\Omega$-plane, the local physics around the singularity is described  by an effective ${\cal N}=2$ supergravity theory with a certain number of hypermultiplets originating from twisted sectors
where all three internal coordinates $X^I$ are twisted (see Appendix  A for details). 
They parametrize the sizes and  shapes of  the compact exceptional cycles at the singularity \cite{Lust:2006zh,Reffert:2006du}.
Twisted sectors where some of the $X^I$ are untwisted preserve larger supersymmetry and contribute non-localised states that are irrelevant to the local physics.

${\cal N}=1$ theories are obtained by the quotient of the orbifold theory by an orientifold involution involving world-sheet parity $\Omega$ combined with a space-time reflection and some additional $\Z_2$ symmetry (eg $(-)^{F_L}$).  
The inclusion of  $\Omega$-planes projects hypermultiplets localised at the singularity onto chiral multiplets describing the sizes of the compact exceptional cycles. 
Fixed points of the reflection define an orientifold plane inverting the orientations of both closed and open strings (to be described next). 
We denote the orientifold action generically by  $\Omega_{\epsilon}$ with  $\epsilon=(\epsilon_0,\epsilon_I)$ four signs satisfying $\prod_{I=1}^3 \epsilon_I=-1$.
These specify the orientation and the charge of the $\Omega$-plane.  
In particular
\bea
 \Omega 3^{\pm} : \qquad && (\pm{\,-}{\,-}{\,-})\nn\\
 \Omega 7^{\pm}: \qquad && (\mp{\,+}{\,+}{\,-})
\eea
represent an $\Omega3$, an $\Omega7$ plane along the (1~2)-planes and so on. The $\epsilon_0=\pm$ sign specifies the Sp/SO projection, with $+$ conventionally taken for the Sp-projection on D-brane stacks coincident with a given $\Omega$-plane.
In a dimer description of the orientifold \cite{Franco:2007ii}, these signs specify the charges of the orientifolds at the four fixed points of the quotiented dimer.

\subsection{Quiver gauge theories}

Next, we consider the inclusion of D-branes at the  $\C^3/\Z_n$ orientifold singularity.  
Compatibly with the ${\cal N}=1$ supersymmetry preserved by the $\Omega$-planes, we consider the insertion of $N$ `fractional' D3-branes as well as $M$ `flavour' D7-branes passing through the singularity and extending along four non-compact directions inside  $\C^3/\Z_n$. 
The dynamics of D7-D7 open strings is irrelevant for the local physics. On the other hand open strings connecting D3 and D7  are localized  at the singularity and provide fundamental matter.  For definiteness we will consider D7 wrapped along the complex planes $I=1,2$, \ie along the non-compact divisor $X^3=0$. One should keep in mind that additional D7 branes wrapped along the non-compact divisor $X^1=0$ or $X^2=0$, or superpositions thereof, can be considered. 
         
To find the field content of the unoriented quiver theory at the singularity we proceed in two steps. Starting from the ${\cal N}=4$ theory living on the D3-branes in flat space-time, we first perform the orbifold projection to an oriented quiver theory with flavour and then perform the unoriented projection to an unoriented quiver theory with flavour. 

In the ${\cal N}=1$ language the ${\cal N}=4$ theory is given by a vector multiplet and three chiral multiplets all in the adjoint of U$(N)$.   In flat space-time D3-D7 open strings contribute $2M$ chiral multiplets ($M$ hypermultiplets) rotated by a U$(M)$ flavour group. 
We denote by $\bf V$ and $\bf C$ a vector and a chiral multiplet of ${\cal N}=1$ supersymmetry, respectively. 
One can then write the field content in the ${\cal N}=1$ language as
\be
\label{n4}
{\cal H}_{\rm flat} =({\bf V}+3 {\bf C})\,  \fund ~ \overline{\fund}   +{\bf C}\, (  \bar{\bf M} \times  \fund  +
 \mathbf M \times \overline{\fund}). 
\ee
Here and in the following we denote by $\fund$($\overline{\fund}$) the (anti)fundamental representation of a gauge group and by its dimension ${\bf M}$ $({\bf \bar M})$ the (anti)fundamental representations of the flavour group. 
The orbifold group breaks  the gauge  and flavour  groups down to $\prod_a {\rm U}(N_a)$ and  $\prod_a  {\rm U}(M_a)$
respectively. Here $N_a$ and $M_a$ denote the number of D3 and D7 branes  transforming in the $a$-representation of $\Z_n$ with $a=0,1\ldots n-1$. Explicitly, the action of the orbifold group generator on Chan-Paton indices 
breaks the fundamental representations of U($N$) and U($M$) 
according to (see Appendix for details)
 \bea
 \label{zorb0}
\Theta : \qquad   \fund  \to \oplus_a   w^a \, \fund_a \,,
\qquad~
{\bf M}  \to \oplus_a   w^a \, {\mathbf  M_a }\,,
\eea
 where we denote by $\fund_a$ and ${\bf M_a}$  the fundamental representations of $ {\rm U}(N_a)$ and $ {\rm U}(M_a)$
 respectively.  In addition, the spacetime action of $\Theta$   on the  field components reads 
\bea
\label{zorb}
\Theta: && \qquad {\bf V}\to {\bf V} ,\qquad  {\bf C}^I\to w^{a_I } {\bf C}^I  ,
\qquad  {\bf C}^{\dot a} \to w^{ s }  {\bf C}^{\dot a} ,
\eea
where by ${\bf C}^I$ and ${\bf C}^{\dot a}$ ($\dot a=1,2$) we denote the chiral multiplets  coming from D3-D3 and D3-D7 strings respectively.
The former transforms in the fundamental of the $ {\rm SU}(3)$ rotation group of $\C^3$ while the latter as a chiral spinor of the rotation group of the $\C^2$ along the D7.    
A consistent orbifold group action on D3-D7 fields requires $s=\frac{a_1+a_2}{2}\in \Z$. For $n$ odd this is not a restriction since one can always redefine $a_I$ by adding $n$.

Combining (\ref{zorb0}) and (\ref{zorb}) and keeping  invariant components in (\ref{n4}) one finds the field content of the oriented quiver gauge theory with flavour 
\be
\label{orb}
{\cal H}_{\rm orbifold}= \sum_{a=0}^{n-1}  
   \left( {\bf V}\,  \fund_a  \, \overline{\fund}_a  
   +   {\bf C} \left[ \sum_{I=1}^3  \, (\fund_a ,\overline{\fund}_{a+ a_I}) +  \mathbf M_a \, \overline{\fund}_{a+s}+ \bar {\bf M}_{a+s}   \, {\fund}_{a}
  \right] \right).   
\ee
More precisely, states in the vector multiplets will be given by $N\times N$ block diagonal matrices, D3-D3 chiral 
multiplets $\Phi^I$ by   $N\times N$ matrices with non trivial components for the $N_a\times N_{a+a_I}$ blocks,
D3-D7 chiral fields $Q$ by  $N\times M$ matrices with $N_a\times M_{a+s}$ non-trivial blocks and 
D7-D3 fields $\tilde{Q}$ by  $M\times N$ matrices with $M_a\times N_{a+s}$ non-trivial blocks. 
Here and henceforth all subscripts will be always understood mod $n$. 
The superpotential is cubic and follows directly from that in flat spacetime
\be
W_{\rm pert}=   {\rm Tr } \left(g\, \Phi^1[\Phi^2,\Phi^3] +h_1\,   \Phi^3 Q\tilde{Q} 
+h_2\,   Q\langle  \Phi^3_{77} \rangle \tilde{Q} \right), \label{wpert}
\ee
after replacing the matrices by their orbifold invariant block form.
The last term, involving the vev of some of the non-dynamical D7-D7 fields can be viewed as a mass terms  in the low energy effective action.
The dimensionless constants $g,h_1,h_2$  measure the strength of the various interactions.

In the absence of D7-branes, tadpole/anomaly cancellation requires $N_a=N_b$ for any $a$ and $b$, corresponding to $N$ copies of the `regular' representation of $\Z_n$. The resulting quiver theory is superconformal in the IR, where anomalous U$(1)$'s decouple or become global (baryonic) symmetries.
The mesonic branch of the moduli space is ${\rm Symm}_N (\rm CY)$ \footnote{This is almost self-evident for ${\cal N}=1$, since $n$ `fractional' D3-branes combine into a `bulk'/regular brane that can wander in CY. A proof for ${\cal N}>1$ remains elusive.}.
The near-horizon geometry is ${\rm AdS}_5\times \mathrm S^5/\Z_n$. Including D7-branes generically spoils superconformal invariance but makes tadpole/anomaly cancellation easier to achieve even without $\Omega$-planes. In particular one can embed the (SUSY) standard model in a flavoured $\Z_3$ quiver \cite{Cicoli:2012vw, Cicoli:2013mpa, Cicoli:2013zha,Dolan:2011qu}.

\begin{figure}
\centering
\raisebox{-0.5\height}{\includegraphics[scale=0.8]{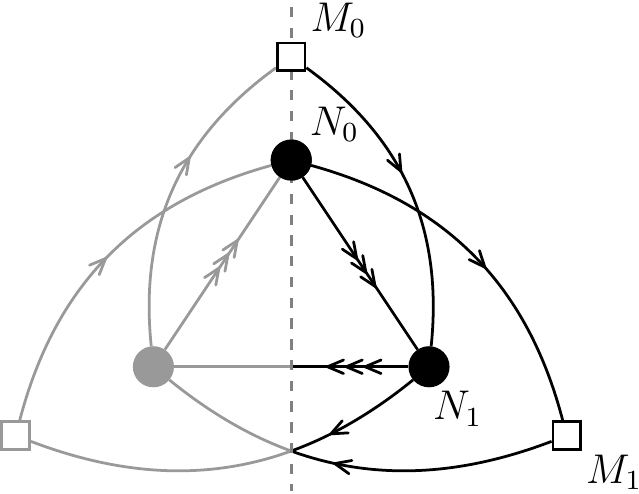}}
\hspace*{-1em}
\raisebox{-0.5\height}{\includegraphics[scale=0.8]{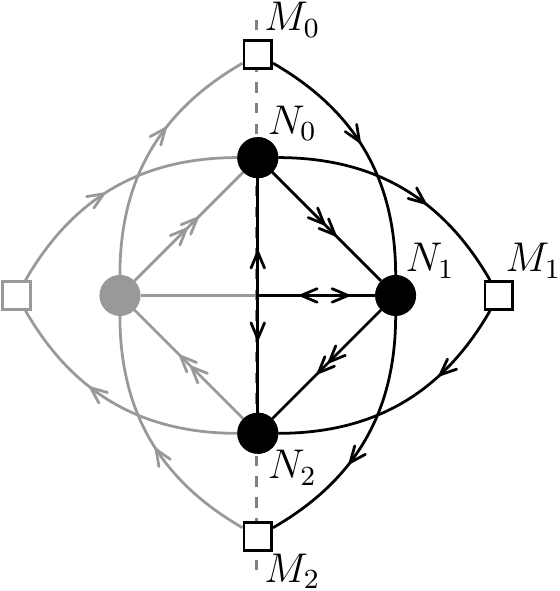}}
\hspace*{1em}
\raisebox{-0.5\height}{\includegraphics[scale=0.8]{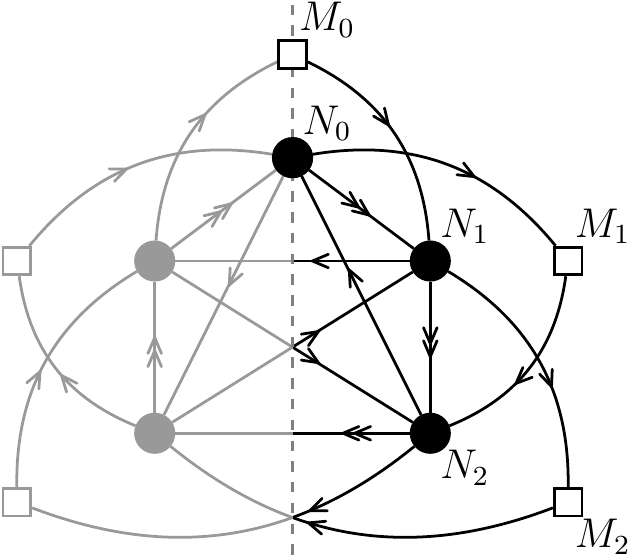}}
\caption[the $\Cint^3/\Zint_3$ and $\Cint^3/\Zint_5$ orientifold theories]
{the $\Cint^3/\Zint_3$, $\C^3/\Z_4$ and $\Cint^3/\Zint_5$ orientifold theories for $a_I=(1,1,-2)$.}
\label{figz3}
\end{figure}

\begin{figure}
\centering
\raisebox{-0.5\height}{\includegraphics[scale=0.8]{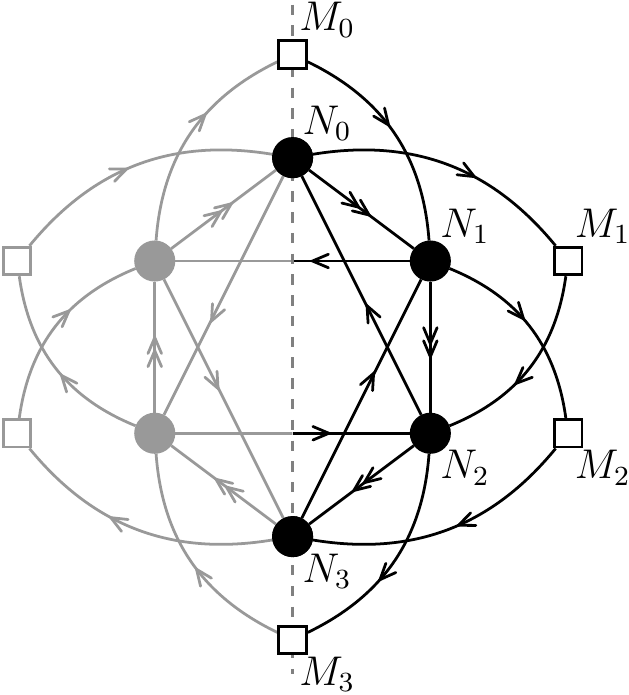}}
\hspace*{2em}
\raisebox{-0.5\height}{\includegraphics[scale=0.8]{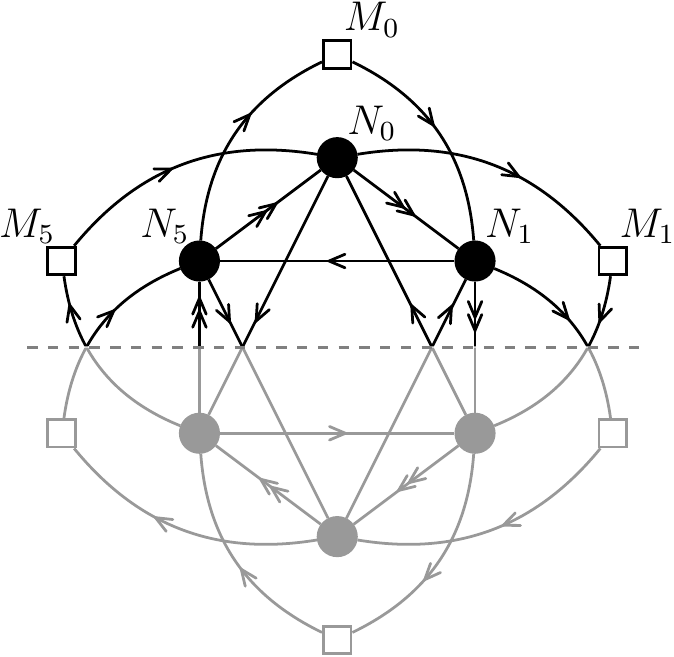}}
\caption[the $\mathbb C^3/\mathbb Z_6$ orientifold theories]
{The $\Cint^3/\Zint_6$ theory, $a_I=(1,1,-2)$, with two different orientifolds, defined in \eqref{omega3} and \eqref{omega3hat} respectively.}
\label{figz6}
\end{figure}

\begin{figure}
\centering
\raisebox{-0.5\height}{\includegraphics[scale=0.8]{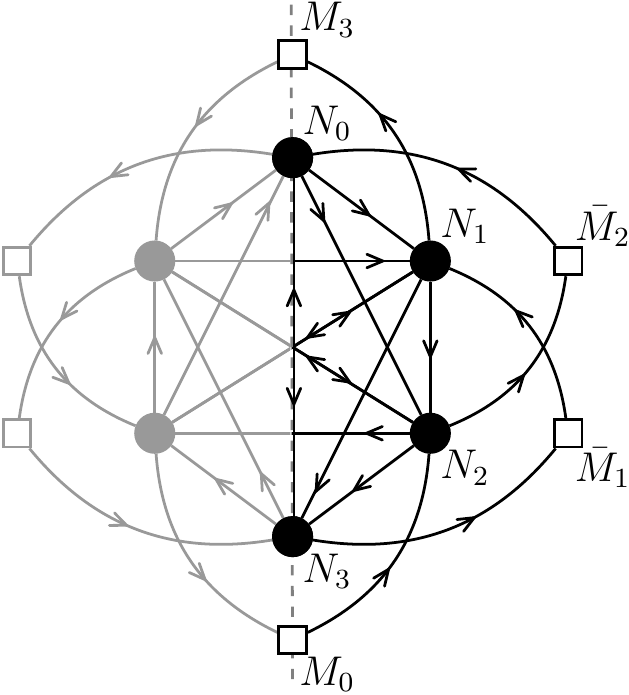}}
\hspace*{2em}
\raisebox{-0.5\height}{\includegraphics[scale=0.8]{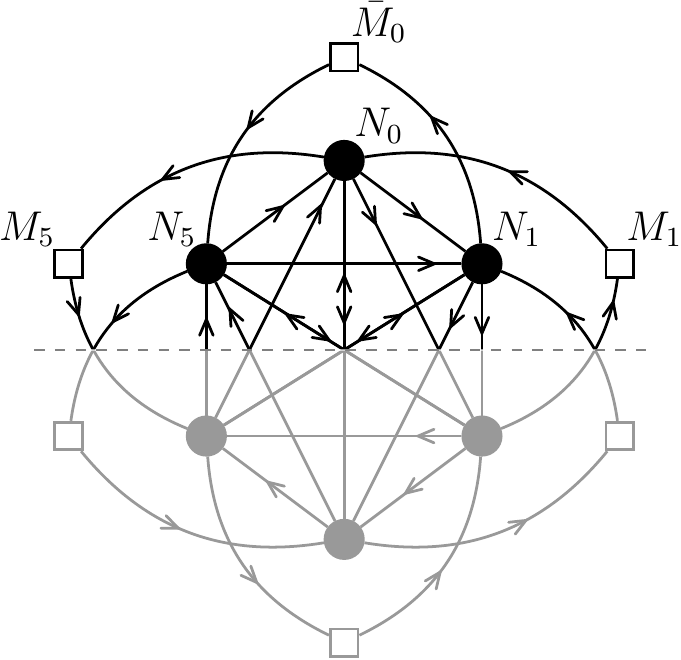}}
\caption[the $\Cint^3/\Zint_6'$ theory with two different orientifolds]
{the $\Cint^3/\Zint_6'$ theory ($a_I=(1,3,2)$) with the two different orientifolds  $\Omega$ and $\hat{\Omega}$.}
\label{figz6prime}
\end{figure}

Let us consider the unoriented projection that  identifies ingoing open strings ending on a brane with the outgoing open strings starting from the image brane transforming in the complex conjugate representation
\bea
\label{omega3}
\Omega_\epsilon: \qquad    \fund_{a}  \leftrightarrow  \overline{\fund}_{n-a} \,,
\qquad ~~~~~~~~~~~~
 \mathbf M_{a}  \leftrightarrow  \bar{\bf M}_{n-a}\,,
\eea
Strings connecting a brane and its image  are projected onto symmetric and antisymmetric representations according to the signs $(\epsilon_0,\epsilon_I)$ specifying the orientifold.
Keeping invariant components from (\ref{orb}) under (\ref{omega3}) one finds
\bea
\label{orient}
{\cal H}_{\rm orientifold} &=& {\bf V}   \left(  \sum_{a=0,{\frac n 2}} \fund^2_{a,\epsilon_0} 
+ \sum_{a=1}^{p}\, \fund_{a} \, \overline{\fund}_a 
 \right) +  {\bf C} \sum_{a=0}^{p}  
   \left(    \bar{\bf M}_{a+s } \,\fund_a +\mathbf  M_a\, \overline{\fund}_{a+s}      
   \right) +\nn\\
&& +{\bf C}  \sum_I \sum_{a=0}^{n-1}  \left\{   
\begin{array}{cc}
  \ft12 (\fund_a ,\overline{\fund}_{a+ a_I})      &   a\neq -a-a_I \\
 \fund_{a,-\epsilon_0\epsilon_I}^2   &   a= -a-a_I \\
\end{array}
\right.
\eea
with $p=[{n-1\over 2}]$  and $\fund^2_{a,\pm}$ denoting the symmetric and antisymmetric representations of  the gauge group at node $a$.
In (\ref{orient})   the identifications $\fund_{a}= \overline{\fund}_{n-a}$ and $ \mathbf M_{a} 
 =\bar{\bf M}_{n-a}$ are understood.  
In particular, one can check that bifundamentals in the last line appear always twice leading to integer multiplicities as expected.  
Examples of unoriented quiver diagrams with flavour are displayed in Figures \ref{figz3}, \ref{figz6} \ref{figz6prime} and \ref{Z5prime}.
The spectrum for $n=3,4,5,6$ and $\epsilon_0 = -1$ is displayed in Table \ref{tab:orientifolds}.\\
For even order orbifold groups $n=2k$ it is also possible to choose another unoriented projection 
\bea
\label{omega3hat}
\hat{\Omega}_\epsilon: \qquad    \overline{\fund}_{a}  \leftrightarrow   \fund_{\frac{n}{2}-a} \,,
\qquad ~~~~~~~~~~~~
 \bar{\bf M}_{a} \leftrightarrow  \mathbf M_{\frac{n}{2}-a}\,,
\eea
which corresponds to an orientifold identifying the node $0$ with the node $n/2$. In Table \ref{tab:orientifolds} we focus on the first new example, the ${\mathbb{Z}}_6$ orbifold with this second orientifold projection.
The corresponding unoriented quiver diagram is in Figure \ref{figz6} on the right. The cases with $n$ multiple of four are equivalent to the previous orientifold projection (\ref{omega3}).

Note that symplectic groups require an even number of (fractional) branes, and this condition applies both to gauge and flavour groups.
Since consistency requires $\Omega$ planes to act with the opposite projections on D3 and D7, one must for instance pay attention to the fact that a theory with an SO$(N_{0})$ gauge group must have even $M_{0}$, since the associated flavour group is Sp$(M_{0})$.

When $n$ is even, the orbifold group also contains the spatial $\Z_{2}$ involution $\Theta^{n\over 2}$. As a result,
$\Omega \Theta^{n\over 2}$ is also an orientifold involution leading an equivalent orientifold group. 
 This leads to the following identifications 
\be\begin{aligned}\label{omega identifications}
\Omega3^{\pm} &= \Omega7^{\mp}\quad\text{($n$ even)},\cr
\hat\Omega3^{\pm} &= \hat\Omega7^{\pm}\quad\text{($\ft{n}{2}$ odd)}.
\end{aligned}\ee

\begin{table}
\centering
\begin{tabular}{ccl}
\toprule
& Gauge Group & Chiral multiplets \& anomalies \\
\otoprule
\multirow{3}{*}{$\Z_3$} & \multirow{3}{*}{$\mathrm{SO}(N_0)\times \mathrm{U}(N_1)$} 
 & $3 ( \fund,\overline{\fund})  +\sum_I (\cdot,\fund^2_{\epsilon_I} )+{\bf \bar M_1} (\fund,\cdot) $\\
&&$   +{\bf  M_0}\, (\cdot,\overline{\fund})+{\bf\bar{M}_1} (\cdot,\fund)$
\\[.2em]
&& $ M_0=\sum_{I=1}^3(N_1-N_0+4\epsilon_I)+M_1 $ \\
\midrule
\multirow{4}{*}{$\Z_4$} & \multirow{4}{*}{$\mathrm{SO}(N_0)\times \mathrm{U}(N_1)\times \mathrm{SO}(N_2)$}   
 & $2( \fund,\overline{\fund},\cdot) + 2(\cdot, \fund,\fund) +(\fund,\cdot,\fund)+(\cdot,\fund^2_{\epsilon_3},\cdot)$ \\
&& $ +(\cdot,\ov{\fund}^2_{\epsilon_3},\cdot)
 +{\bf \bar M_{1} }(\fund,\cdot,\cdot)+{\bf \bar M_{2} }(\cdot,\fund,\cdot)$ \\
&& $+{\bf M_{0}}(\cdot,\overline{\fund},\cdot)  + {\bf M_{1}} (\cdot,\cdot,{\fund})$ \\[.2em] 
&& $M_0=-2N_0 +2N_2   +M_2$ \\
\midrule
\multirow{5}{*}{$\Z_5$} & \multirow{5}{*}{$\mathrm{SO}(N_0)\times \mathrm{U}(N_1)\times \mathrm{U}(N_2)$}   
 & $2( \fund,\overline{\fund},\cdot) + 2(\cdot, \fund,\overline{\fund}) +(\fund,\cdot,\fund) 
   +(\cdot,\overline{\fund},\overline{\fund}) +(\cdot,\fund^2_{\epsilon_3},\cdot)$ \\
&& $+(\cdot,\cdot,\fund^2_{\epsilon_1}) +(\cdot,\cdot,\fund^2_{\epsilon_2}) 
 +{\bf \bar M_{1} }(\fund,\cdot,\cdot)+{\bf \bar M_{2} }(\cdot,\fund,\cdot)$ \\
&& $+{\bf M_{0}}(\cdot,\overline{\fund},\cdot)  + {\bf M_{1}} (\cdot,\cdot,\overline{\fund}) +{\bf M_{2}}(\cdot,\cdot,\fund)$ \\[.2em] 
&& $ M_0 =-2N_0 +N_1 +N_2+4\epsilon_3 +M_2$ \\
&& $M_1 = N_0 -3N_1 +2N_2+4(\epsilon_1+\epsilon_2) +M_2  $ \\
\midrule
\multirow{6}{*}{$\Z_6$} 
&& $2(\fund,\overline{\fund},\cdot,\cdot) + 2(\cdot,\fund,\overline{\fund},\cdot) 
   +2(\cdot,\cdot,\fund,\fund) +(\fund,\cdot,\fund,\cdot)$ \\
&& $+(\cdot,\overline{\fund},\cdot,\fund) +(\cdot,\fund^2_{\epsilon_3},\cdot,\cdot) 
   +(\cdot,\cdot,\ov{\fund}^2_{\epsilon_3},\cdot)$ \\
&$\mathrm{SO}(N_0)\times \mathrm{U}(N_1)$ & $+{\bf M_{0}}(\cdot,\overline{\fund},\cdot,\cdot) +{\bf\bar M_{1}}(\fund,\cdot,\cdot,\cdot)+{\bf\bar M_{2} }(\cdot,\fund,\cdot,\cdot)$ \\
&$\quad{}\times \mathrm{U}(N_2)\times \mathrm{SO}(N_3)$ & $ + {\bf M_{1}} (\cdot,\cdot,\overline{\fund},\cdot) +{\bf M_{2}}(\cdot,\cdot,\cdot,\fund)+{\bf\bar M_3}(\cdot,\cdot,\fund,\cdot)$ \\[.2em] 
&& $M_0=-2N_0 +N_1 +2N_2 -N_3   +M_2 +4\epsilon_3$ \\
&& $ M_1 = N_0 -2N_1 -N_2 +2N_3   +M_3 -4\epsilon_3$ \\
\midrule
\multirow{6}{*}{${\Z}_6,\,\hat\Omega$} & \multirow{6}{*}{$\mathrm{U}(N_0)\times \mathrm{U}(N_1)\times \mathrm{U}(N_5)$}      
 & $2(\fund,\ov{\fund},\cdot)+2(\ov{\fund},{\cdot},{\fund})+(\ov{\fund},\ov{\fund},\cdot)+({\fund},\cdot,\fund)+ (\cdot,\fund,\ov{\fund}) $ \\
&& $+2(\cdot,\fund^2_{\epsilon_i},\cdot)+2(\cdot,\cdot,\ov{\fund}^2_{\epsilon_i})+\bar{\bf M}_0(\cdot,\cdot,{\fund})+{\bf M}_0(\cdot,\ov{\fund},\cdot)$\\
&&$+\bar{\bf M}_1({\fund},\cdot,\cdot)
+{\bf M}_1(\cdot,{\fund},\cdot)+\bar{\bf M}_5(\cdot,\cdot,\ov{\fund})
+{\bf M}_5(\ov{\fund},\cdot,\cdot) $ \\[.2em]
&& $M_1=3N_0-2N_1-N_5 -4(\epsilon_1+\epsilon_2)+M_0$ \\
&& $M_5=3N_0-N_1-2N_5 -4(\epsilon_1+\epsilon_2)+M_0 $ \\
\bottomrule
\end{tabular}
\caption{Matter content for some $\Cint^3/\Zint_n$ orientifold theories, with $a_I=(1,1,-2)$ and $\epsilon_0=-1$.
The field content for $\Omega$ projections of Sp$(N)$ type (corresponding to $\epsilon_0=1$) follows by flipping all antisymmetric into symmetric representations and vice-versa, i.e.  $\epsilon_I \to -\epsilon_I$. 
The constraints on $M_i$ come from the tadpole cancellation conditions.}
\label{tab:orientifolds}
\end{table}

\subsection{Tadpoles and anomalies}

For generic choices of $N_a$ and $M_a$, the unoriented quiver gauge theories obtained in the last section are chiral and therefore potentially anomalous. Sp and SO gauge groups are free of anomalies since, barring spinorial representations that are not realised in perturbative open string contexts, all representations are self-conjugate. 
For U$(N)$ gauge groups the anomaly is computed by the formula
\be
\mathcal I_{U(N)}=\Delta n_F +\Delta n_A (N-4) +\Delta n_S (N+4),
\ee  
with $\Delta n_F$, $\Delta n_A$ and $\Delta n_S$ the differences between the number of chiral and anti-chiral ${\cal N}=1$ multiplets in the fundamental, symmetric and antisymmetric representations respectively. Higher rank (anti-)symmetric tensors are not realised in perturbative open string contexts.
Taking into account the field content of the unoriented quiver gauge theory one finds
\be
{\mathcal I}_a=\mathcal I_{{\rm U}(N_a)} = \sum_{b=0}^{n-1} ( I_{ab} \,N_b+ J_{ab}\, M_b) +4 \epsilon_0 K_{a}  
 \label{tadfin}
\ee
with
\bea
I_{ab} &=& \sum_{I=1}^3 ( \delta_{a,b-a_I} -\delta_{a,b+a_I}) \nn\\
J_{ab} &=&  \delta_{a,b-s} -\delta_{a,b+s }\nn\\
K_a &=& \sum_{I=1}^3  \epsilon_I ( \delta_{2a,a_I} -\delta_{2a,-a_I}  ) \label{inters}
\eea
codifying the ``intersection numbers'' of the exceptional cycles at the singularity. 
More concretely, $I_{ab}$ counts the number of times D3 branes of type ``$a$" and ``$b$"
intersect, $J_{ab}$ the intersections of D3$_a$ and D7$_b$ branes and $K_a$ the intersections
of a D3$_a$ brane and its image D3$_a'$ under the orientifold action. This can be read off directly from the quiver diagram counting the
number of arrows connecting the various nodes with plus or minus signs depending on the direction of the arrow. 
We notice that $I_{ab}$ and $J_{ab}$ are anti-symmetric matrices while $K_a=-K_{n-a}$. 
Explicitly for $a_I=(1,1,-2)$ the non-trivial components are
\begin{align}
\Z_3\quad &   I_{a,a+1}=-I_{a+1,a}=3,  \quad  J_{a,a+1}=-J_{a+1,a}=1, \quad K_2=-K_1=\sum_I\epsilon_I \,,
\\[.5em]
\Z_{n\neq 3}\quad &     I_{a,a+1}=-I_{a+1,a}=2, \quad    I_{a+2,a}=-I_{a,a+2}=1,
\quad J_{a,a+1}=-J_{a+1,a}=1, \cr
&  K_{ {n+ 1\over 2} }\hspace*{1pt}=-K_{n-1\over 2}= (\epsilon_1+\epsilon_2), \quad    K_{ {n- 2\over 2} }=-K_{ {n+2\over 2} }=-K_1=K_{n-1}= \epsilon_3.
\end{align}

For the even $n$ cases with orientifold projection $\hat{\Omega}_{\epsilon}$, defined in (\ref{omega3hat}), the previous expression for $K_a$ is replaced by
\be
\hat{K}_a = \sum_{I=1}^3  \epsilon_I (\delta_{2a,a_I+\frac{n}{2}} -\delta_{2a,\frac{n}{2}-a_I}  ),
\ee
with the same meaning of intersections between a D3$_a$ brane and its image. In the following sections we will mainly focus on the cases with the $\Omega_{\epsilon}$ projection defined in (\ref{omega3}).\\
We remark that equation (\ref{tadfin}) can be thought of as the components of the vector equation
\be
 N_b\, \pi_{D3b}+M_b\, \pi_{D7b} +4 \epsilon_0\, \pi_{O}=0   \label{tadvec}
\ee  
with $\pi_{D3b}$, $\pi_{D7b}$, $\pi_{O}$ the cycles wrapped by the ${\rm D3}_b$, ${\rm D7}_b$ and $\Omega$-planes respectively.
Equation (\ref{tadfin}) follows from (\ref{tadvec}) after multiplying it by $\pi_a$ and identifying
\be
I_{ab}=\pi_{D3a} \circ \pi_{D3b},\qquad  J_{ab}=\pi_{D3a} \circ \pi_{D7b},\qquad 
K_{a}=\pi_{D3a} \circ \pi_{O}   \label{prodos}.
\ee
We would like to stress that the above `intersection numbers' are completely coded in the various contributions to the one-loop Klein bottle, Annulus and Moebius strip amplitudes.
The interested reader can find all the details in the Appendix.
As already observed long time ago \cite{Bianchi:2000de,Uranga:2000xp}, chiral anomalies are associated to tadpoles of twisted RR fields localized at the singularity and thus belonging to sectors with non-vanishing Witten index, \ie giving rise to an ${\cal N} =1$ (chiral) spectrum.
Tadpoles of RR fields belonging to the untwisted sector or to twisted sectors with vanishing Witten index i.e. giving rise to an ${\cal N} =4,2$ (chiral) spectrum, do not contribute to chiral anomalies in $D=4$ and can thus be discarded in the low-energy dynamics of the local unoriented quiver gauge theory.
Additional constraints arise when one looks for a global embedding of these models. We will not address these important issues here since we are focussing on the local models.  For recent work see \cite{Cicoli:2013mpa}.

\subsection{ $\C^3/\prod_i \Z_{n_i}$-singularities}

Although in explicit examples we have mostly focused on the $\Z_n$ case with $a^I=(1,1,-2)$, formulae in the previous section apply  to the general case $a^I\neq (1,1,-2)$ and to the case of type IIB orientifolds 
on $\C^3/\prod_i \Z_{n_i}$. The singularity is now codified in the choice of the vectors $\vec a_{I} =\{  a^{(i)}_{I } \}$
satisfying
\be
\sum_{I=1}^3 a^{(i)}_{I }=0  \quad ({\rm mod} \:  n_i)
\ee
for each $i$ separately.
The spectrum, anomalies and tadpoles are given by the same formulae as before with intersection numbers $I_{\vec a \vec b}$,
$J_{\vec a \vec b}$,  $K_{\vec a}$, where we define $\vec a = (a^{(1)},a^{(2)},\ldots)\in\Z_{n_1}\times\Z_{n_2}\times\ldots$ The resulting intersection matrices are the tensor product of those of each single $\Z_{n_i}$ factor.
As an example, let us consider $\C^3/\Z_2\times \Z_3$ with the following actions on $\C^3$:
\be
a^{(1)}_I =(1,-1,0)  , \qquad    a^{(2)}_I =(0,-1,1).
\ee
The nodes of the quiver are labeled by $\vec a=(a^{(1)},a^{(2)})$ with $a^{(1)}=0,1$ and $a^{(2)}=0,1,2$, so we have six nodes. One can then see that this orbifold action is precisely identical to $\C^3/\Z_6$ with $ a^{(1)}_I =(1,3,2)$.

At the cost of being pedantic, there is a single fixed point, the origin, in $\C^3/\prod_i \Z_{n_i}$ and closed string (chiral) amplitudes with ${\cal N} =4,2$ supersymmetry do not contribute to tadpole, since the corresponding (un)twisted fields are not localised at the singularity but de-localised along non-compact cycles.

\section{Conformal theories}\label{sec:conformal theories}
\label{sconformal}

Although generically the presence of flavour D7-branes and $\Omega$-planes tends to spoil superconformal invariance, judicious choices of the numbers and types of D7's may lead to $\mathcal N=1$ superconformal quiver gauge theories, thus opening up a completely new class of gauge theories of this kind that are amenable to a reliable description in terms of open strings. 

The prototype is the class of $\mathcal N=2$ superconformal gauge theories arising from $N$ D3's in the presence of 4 D7's and an $\Omega 7^-$ plane \cite{Sen:1996vd,Banks:1996nj,Aharony:1998xz}. The resulting gauge group is Sp$(2N)$, the flavour symmetry is SO$(8)$ acting on the 8 half hypermultiplets in the fundamental representation. In addition there is a flavour singlet hypermultiplet transforming in 
the anti-symmetric representation. The one-loop $\beta$-function of the Sp$(2N)$ gauge theory vanishes and
since for a theory with ${\cal N}=2$ supersymmetry no anomalous dimensions are generated for hyper-multiplets,
one can safely argue that the theory is (super)conformal.  Here we consider  ${\cal N}=1$ theories obtained as orbifold projections
of ${\cal N}=2$ theories, so it is reasonable to believe that again anomalous dimensions for the fundamental fields be not
generated.  Indeed, superpotential interactions are always cubic so chiral fields come with their naive dimension one, as long as vev's of the non-dynamical D7-D7 fields appearing in (\ref{wpert}) vanish.
To look for a superconformal theory one can then scan for models with vanishing one-loop $\beta$-function\footnote{We remark that these arguments can be easily adapted even to non-supersymmetric models of the class \cite{Angelantonj:2000kh,Bianchi:2000vb} where each individual sector preserves some 
 supersymmetry and therefore no tadpoles for the dilation and other NS-NS fields are generated.}.

With this proviso, the one-loop $\beta$ function for a general ${\cal N}=1$ gauge theory  is
\be
\beta=\ft12(3  \ell({\bf Adj}) - \sum_{C} \ell({\bf R_C})  )  ,
\ee
with the sum running over the chiral multiplets and $\ell({\bf R})$ denoting the index of the representation ${\bf R}$. In our
conventions
\be
  \ell(\fund\, \overline{\fund} )=2N, \qquad \ell(\fund^2_\epsilon)=\ell(\overline{\fund}^2_\epsilon)=N+2 \epsilon,  \qquad \ell(\fund)=\ell(\overline{\fund})=1.
\label{lls}
\ee
 For the  quiver gauge theories under consideration one finds
\bea
\label{beta dimf delta}
{\beta_a} &=&
\left\{\begin{array}{ll}
3 N_a+\epsilon_0 K^+_a-\frac{1}{2}\left(I^+_{ab}N_b+J^+_{ab}M_b\right)
& \quad{\rm (SU)}
\\[.2em]
\frac{3}{2}N_a+3\epsilon_0+\frac12\epsilon_0 K^+_a-\frac{1}{4}\left(I^+_{ab}N_b+J^+_{ab}M_b\right)&\quad{\rm (SO/Sp)}
\end{array}\right. 
\eea
in terms of
\be
\label{inters symm}
I^+_{ab} = \sum_{I=1}^3 ( \delta_{a,b-a_I} +\delta_{a,b+a_I}), \qquad
J^+_{ab} =  \delta_{a,b-s} +\delta_{a,b+s },\qquad
K^+_a = \sum_{I=1}^3  \epsilon_I ( \delta_{2a,a_I} +\delta_{2a,-a_I}  ),
\ee
counting the number of arrows (independently of their orientations) in the quiver diagram connecting D3-D3, D3-D7 and D3-D3$'$ branes respectively.
Using (\ref{beta dimf delta}) it is indeed straightforward to impose the vanishing of the one-loop beta function coefficients, obviously together with the tadpole cancellation conditions.

\begin{table}[h!]
\centering
\begin{tabular}{llll}
\toprule
    & $\Omega$ plane & Conformal theories  & Flavour branes 
\\ \otoprule
$\Z_3$ 
  & $\Omega7^-$ 
  & Sp$(N)\times{\rm U}(N+1)$
  & $M_0=2,\ M_1=3$
\\[.5em]
$\Z_4$ 
  & $\Omega3^+/\Omega7^-$ 
  & Sp$(N)^2\times{\rm U}(N+3-p)$
  & $M_0=M_2=4-2p,\ M_1=2p$,~ $p=0,1,2$
\\[.5em] 
$\Z_5'$ 
  & $\Omega7^-$ 
  & Sp$(N)\times{\rm U}(N+1)^2$
  & $M_0=0,\ M_1=1,\ M_2=3$
\\[.5em] 
$\Z_6'$
  & $\Omega3^+/\Omega7^-$
  & Sp$(N)^2\times{\rm U}(N+3)^2$
  & $M_0=M_3=4,\ M_1=M_2=0$
\\[.5em] 
${\Z}_6'$
  & $\hat{\Omega}3^-/\hat{\Omega}7^-$
  & U$(N)\times{\rm U}(N+1)^2$
  & $M_0=4,\ M_1=M_5=0$
\\ \bottomrule
\end{tabular}
\caption{Examples of superconformal unoriented quiver gauge theories.}
\label{tab:full cft}
\end{table}

\begin{figure}[t]
\centering
\raisebox{-0.5\height}{\includegraphics[scale=0.8]{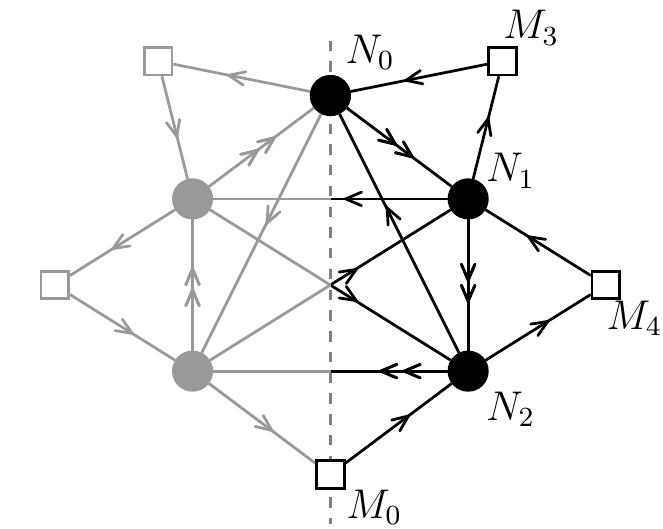}}
\caption{the $\C^3/\Z_5$ theory with $a_I=(1,3,1)$.}
\label{Z5prime}
\label{figz5SCFT}
\end{figure}

We distinguish between two classes of solutions: theories where $\beta_a=0$ for all nodes $a$, and theories which have non-conformal but empty nodes ($\beta_a\neq0$ for $ N_a=0$).
For $\C^3/\Z_n$ models with $n=3,\ldots6$ and $\Omega3$ or $\Omega7$ planes, %
we have found seven new conformal models, whose properties are summarized in Table \ref{tab:full cft}.
The $\C^3/\Z_5'$ case corresponds to $a_I = (1,3,1)$, so that the structure of the flavour representations is changed since in this case $s=(a_1+a_2)/2 = 2$. 
Its quiver diagram is depicted in Figure~\ref{Z5prime}

If we choose one node of the quiver to be empty, $N_a=0$, and relax the associated constraint $\beta_a=0$ for conformal invariance, it turns out to be much easier to find new conformal models.
For brevity, we only provide few examples in table \ref{tab:cft with empty node} for the $\Z_{3}$ orbifold with $\Omega3$ and $\Omega7$ planes.
One can easily find many more models for other orbifolds and/or allowing for more than one non-conformal empty node.
Looking at Tables \ref{tab:full cft} and \ref{tab:cft with empty node}, we see that all solutions require the presence of (fractional) D7 flavour branes to compensate for the superconformal breaking $\Omega$-plane contribution.

It is particularly interesting to note that all models in Table~\ref{tab:full cft} can be seen as $\mathcal N=1$ truncations of the $\mathcal N=2$ Sp$(N)$ superconformal theories discussed in \cite{Sen:1996vd,Banks:1996nj,Aharony:1998xz}.
Indeed, not only all these models have a $\Omega7^-$ plane, but also the total number of D7 branes is always 4, reproducing the (local) setup of the F-theory solution of \cite{Sen:1996vd}.\footnote{One must keep into account that D7 branes on top of the orientifold are counted twice.}

\begin{table}[h!]
\centering
\begin{tabular}{llll}
\toprule
  &$\Omega$ plane & Conformal theories  & Flavour branes 
\\ \otoprule
$\Z_3$ &
$\Omega3^+$ 
  & Sp($N$)
  & $M_0=18,\ M_1=3N+6$
\\[.3em] 
&$\Omega7^-$ 
  & Sp$(N)$
  & $M_0=2,\ M_1=3N+6$
\\[.3em]
&$\Omega3^-$   
  & SO$(0)\times{\rm U}(N)$
  & $M_0=3(N-1),\ M_1=9\quad (N\ \rm odd)$
\\[.3em]
&$\Omega7^-$
  & Sp$(0)\times {\rm U}(N)$
  & $M_0=3N-1,\ M_1=3$
\\ \bottomrule
\end{tabular}
\caption{Conformal theories found for the $\Z_{3}$ orbifold with one non-conformal empty node.}
\label{tab:cft with empty node}
\end{table}

It would be interesting to study whether these superconformal unoriented quiver theories admit a holographic dual. 
One would expect a gravity dual on AdS$_5\times X$ with $X$ a deformation of the Einstein space $S^5/\Z_n$ 
accounting for the presence of the fractional and flavour branes (see \cite{Karch:2002sh,Ouyang:2003df} for previous works in this direction). In this context, tadpole conditions translate into constraints on the volumes of the various non-trivial cycles (faces of the dimer) of $X$. One can take the complementary attitude and exploit the world-sheet description of the brane system to  study the `holographic'  dual gravity solution of the RG flow triggered by the disk  `dilaton'  tadpoles along the lines of \cite{Leigh:1998hj, Angelantonj:2000kh, Bianchi:2000vb, Bertolini:2000dk,Billo:2011uc,Fucito:2011kb,Billo:2012st}.

\section{Instanton induced superpotentials}
\label{sinstanton}

We now turn our attention to non-perturbative effects generated by D-brane instantons 
in unoriented quiver theories with flavour. As by now customary, we start from the oriented case and then consider the effect of the unoriented projection and the inclusion of flavour branes.

In flat space-time as well as in AdS (near horizon geometry), D-instantons behave as instantons for the ${\cal N} =4 $ SYM on a stack of D3-branes \cite{Billo:2002hm, Bianchi:1998nk, Witten:1998xy}.
In the  quiver gauge theories, just like fractional D3-branes correspond to D5 ad D7-branes wrapping vanishing cycles at the singularity, instantons can be realized in terms of fractional D(-1)-branes, i.e. Euclidean D1 and D3-branes wrapping the same set of vanishing cycles.
The orientifold projection restricts these choices to configurations with zero net D5-brane charge.
Unoriented D-brane instantons  have been considered for their crucial role in generating phenomenologically interesting couplings in the superpotential \cite{Ibanez:2006da,Blumenhagen:2006xt,Argurio:2007vqa,Bianchi:2007wy}. For a recent review see \cite{Bianchi:2009ij, Blumenhagen:2009qh, Bianchi:2012ud} and references therein.  Lately the analysis has been extended to (fluxed) E3-branes in F-theory \cite{Cvetic:2010rq, Grimm:2011dj, Bianchi:2011qh, Bianchi:2012kt}. 

In (unoriented) ${\cal N} =1 $ quiver theories, instanton induced superpotentials $W$ are computed by means of the instanton partition function
\be
S_{W}=\prod_{a=1}^{\left[{n\over 2}+1\right]} \Lambda_a^{k_a \beta_a} \, \int d\mathfrak{M} \,e^{S_{\rm inst}}=\int d^4 x \,d^2\theta\,  W(\Phi),
\ee
with $\mathfrak{M}$ the ADHM moduli space realized in terms of open strings with at least one end on the D-instanton
($ d^4x\, d^2\theta$ is the center of mass super-volume form).
$\Lambda_a$, $\beta_a$, $k_a$ are the scales, beta functions and instanton numbers associated to the gauge group
at node $a$ and $S_{\rm inst}$ is the instanton moduli space action. 

There are two distinct classes of instantons: gauge and exotic instantons. Gauge instantons
are associated to a single D(-1) brane (and its image) occupying a non-empty node of the quiver (\ie wrapping the same vanishing cycle as a physical stack of branes) and
generate  Affleck--Dine--Seiberg like superpotentials. Exotic instantons arise from a single D(-1) brane occupying a Sp empty node and generate polynomial superpotential terms\footnote{The effect of E3 instantons associated to flavour nodes vanishes in the strict non-compact limit but may resurrect when the local unoriented quiver is embedded in a consistent global context.}. 

\subsection{Gauge Instantons}

Let us first consider `gauge' instantons.
The instanton fermionic moduli space can be splitted into two classes according to whether the zero mode corresponds to the gaugino (vector multiplet) or to matter fermions (chiral multiplets). We denote the total number of them for $k=1$ by $n_{\lambda_0}$ and $n_{\psi_0}$ respectively.  Index theorems yield 
\be
n_{\lambda_0}= \ell({\bf Adj}),   \qquad 
n_{\psi_0} = \sum_{C} \ell({\bf R_C}) ,
\ee
with the sum running over the chiral multiplets and $\ell({\bf R})$ given in (\ref{lls}). 
The beta function of the gauge theory is given by 
\bea
\beta=\ft12(3 n_{\lambda_0} - n_{\psi_0}  ).  \label{betann}
\eea
A single instanton can generate a superpotential \`a la Affleck-Dine-Seiberg if
$ n_{\lambda_0}-n_{\psi_0} =2$, like in SQCD with $N_f = N_c -1$. In this case all fermionic zero modes, except for the two  $\theta$'s parametrizing the superspace coordinates, can be soaked by bilinear terms in the fermion zero-modes  arising from Yukawa couplings. Plugging this condition into (\ref{betann}),  one concludes  that a superpotential can be generated if the beta function of the gauge theory satisfies the condition
\be
\beta=\ell({\bf Adj})+1=
\left\{
\begin{array}{cc}
   2N+1 & {\rm U}(N)  \\
  N+3&   {\rm Sp}(N)   \\
  N-1  & {\rm SO}(N)  
\end{array}
\right.\label{betainst}
\ee
The generated superpotential can be written in the form
\be
 W_{\rm gauge}= {\Lambda^{\beta} \over \Phi^{\beta-3}} ,
\ee
where $\Phi^{\beta-3}$ is some gauge and flavour invariant composite operator, whose `refined' expression in terms of the chiral matter super-fields takes into account the exact number of zero-modes of each kind, \ie for $\Z_3$ with no flavour branes and gauge group  SU$(4)$, $\beta = 9$ and
 $\Phi^{\beta -3} = {\rm det}_{3\times 3}\, \epsilon_{u_1..u_4} \phi^{u_1 u_2}_{I} \phi^{u_3 u_4}_{J} $ \cite{Bianchi:2007wy}.

It is now easy to scan table \ref{tab:orientifolds} for unoriented quiver gauge theories with flavour admitting nodes such that the beta function satisfies  (\ref{betainst}). In these cases a superpotential term can be induced by gauge instantons.  In  table \ref{thz345inst} we collect the quiver gauge theories exhibiting   superpotentials of this type.\\
\begin{table}[h!!]
\centering
\begin{tabular}{clll}
\toprule
   & Gauge theories  & Flavour branes 
\\ \otoprule
\multirow{1}{*}{$\Z_3$}  
   & ${\rm Sp}(2p)_*$ 
   & $M_0=4(3-p),\ M_1=2p$
   & $p=0,\ldots,3$ 
\\[.2em]
   & ${\rm Sp}(2p)_*\times {\rm U}(1)$ 
   & $M_0=4(3-p),\ M_1=2p-3$
   & $p=2,3$ 
\\[.2em]
   & ${\rm Sp}(6)_*\times {\rm U}(2)$ 
   & $M_0=M_1=0$
   &  
\\[.2em]
  & ${\rm SO}(0)\times {\rm U}(4)_*$ 
   & $M_0=M_1=0$
   &  
\\
\midrule
\multirow{1}{*}{$\Z_4$}
   &Sx$(N_0)_*$$\times$U($N_1$)$\times$Sx$(N_2)$
   & $M_1=N_0$$-$$N_2$$-$$2N_1$$-$$2(1$$-$$\epsilon_0)$ 
   & $N_0\ge2(1$-$\epsilon_0)$+$2N_1$+$N_2$
\\
   &
   & $M_2=M_0+2N_0-2N_2$
   & 
\\[.3em]
   &Sx$(N_0)$$\times$U($N_1$)$\times$Sx$(N_2)_*$
   & $M_1=N_2$$-$$N_0$$-$$2N_1$$-$$2(1$$-$$\epsilon_0)$ 
\\
&
   & $M_0=M_2+2N_2-2N_0$
   & $N_2\ge 2(1$-$\epsilon_0)$+$2N_1$+$N_0$
\\ \bottomrule
\end{tabular}
\caption{Chiral gauge theories at the $\mathbb{Z}_{n}$, $n=3,4$,  orientifold singularities admitting instanton contributions. The node where the instanton sits is indicated by a $*$. We use the symbol Sx$\,\equiv\,$SO, Sp for $\epsilon_0=-1,\ +1$ respectively. Recall that for $\epsilon_0=-1$ $M_0$ and $M_2$ must be even.}
\label{thz345inst}
\end{table}
For the $\mathbb{Z}_3$ and $\mathbb{Z}_5$ orbifolds the number of solutions is finite. In particular for the $\mathbb{Z}_3$ case these solutions extend the gauge theories ${\rm Sp}(6)_*\times {\rm U}(2)$  and ${\rm U}(4)_*$    found in \cite{Bianchi:2007wy} without D7 branes. The $*$ indicates the gauge group where the instanton sits.

We conclude this section by remarking that instantons may generate different dynamical effects. Indeed for gauge theories with $\beta=\ell({\bf Adj})$ one finds that, like for QCD with $N_f=N_c$, the moduli space can get
deformed  at the scale $\Lambda$ (see for instance \cite{Argurio:2007qk,Bianchi:2009bg}). On the other hand, there may be other non-perturbative effects, that may be related to instantons after Higgsing, leading to dynamical super potentials. In particular
$\beta=\ell({\bf Adj})-1$ is a necessary condition for S-confinement \cite{Csaki:1996sm,Csaki:1996zb,Grinstein:1997zv,Grinstein:1998bu}	, like in QCD with $N_f = N_c + 1$.  
For example, for the $\Z_4$ quiver one can find gauge theories with:

\begin{itemize}

\item{${\rm Sp}(2p)_*\times {\rm U}(0)\times {\rm Sp}(2p)_* $: Two types of instanton superpotentials are generated at each of the two non empty gauge theory nodes with scales $\Lambda_0$ and $\Lambda_2$.}

\item{${\rm Sp}(2p+2)_*\times {\rm U}(N_1)\times {\rm Sp}(2p) $ with $N_1=0,1$: 
 A superpotential is generated  by a gauge instanton at node 0 while the theory S-confines at node 2 .}

\end{itemize}

\subsection{Exotic Instantons}


Exotic instantons originate from a single D(-1) occupying an empty Sp node and carrying an O(1) symmetry.  For this choice the instanton moduli space  contains (besides the two universal fermionic zero modes and the four positions) only fermionic zero modes coming from D(-1)-D3 or D(-1)-D7 strings.   
Assuming that the D(-1) sits in node 0, the number of fermionic zero modes is summarized in the following table
\begin{center}
\begin{tabular}{cccc}
type & modes & $U(N_b)$ & ${\rm dim} \mathfrak{M}_F$\\
{\rm D}(-1)-{\rm D}(-1)&$ x_\mu,\ \theta_\alpha$&$\bullet$ & 2\\
{\rm D}(-1)-{\rm D}3&$\mu^I$&$\fund_{a_I}$& $\sum_I N_{a_I}$\\
{\rm D}(-1)-{\rm D}7&$\mu'$&$  {\bf M}_{s}\times \bullet $ & $M_{s}$
\end{tabular}
\end{center}
  For a D(-1) instanton at node $\ft{n}{2}$, a similar spectrum is found with $a_I \to \ft{n}{2}+a_I$ and $s\to s+\ft{n}{2}$.    
 A non-perturbative superpotential arises whenever it is possible to saturate the integration over the charged moduli $\mu^I,\ \mu'$ and again the superpotential can be written in the form
 \be
 W_{\rm exotic}=\Lambda^\beta \, \Phi^{3-\beta},
 \ee
 with $\beta\leq 3$ the putative beta function of the Sp(0) node
 \be
 \beta=3-\ft12 (\sum_I N_{a_I}+ M_s).
 \ee
 
\subsection*{Examples}

\begin{figure}
\centering
\raisebox{-0.5\height}{\includegraphics[scale=.8]{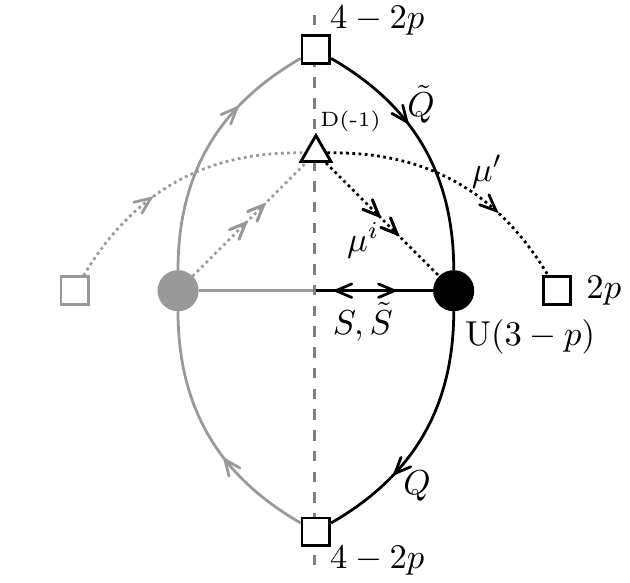}}
\hspace*{3em}
\raisebox{-0.5\height}{\includegraphics[scale=.8]{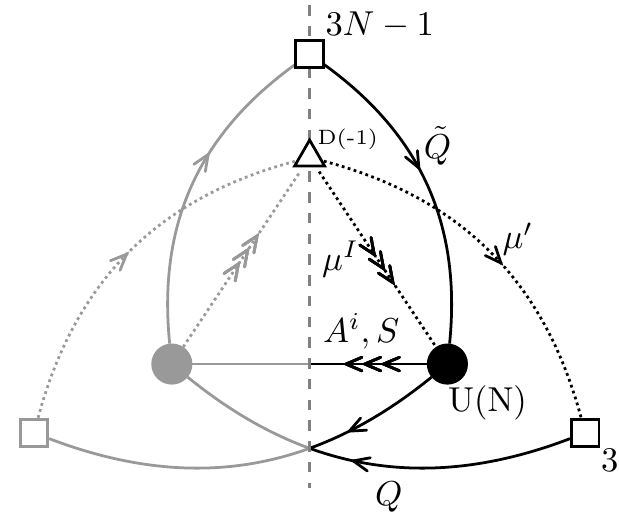}}
\caption[the $\Cint^3/\Zint_3$ and $\Cint^3/\Zint_5$ orientifold theories]
{the $\C^3/\Z_4$ U($3-p$) (left) and $\C^3/\Z_3$ U($N$) (right)  models admitting exotic instanton contributions.}
\label{fig:exoticinstanton}
\end{figure}

In the following, we discuss two examples of instanton induced superpotentials in ${\cal N}=1$ superconformal unoriented quiver gauge theories.

As a first example, consider the U($3-p$) conformal theory that one can obtain from the second row of Table \ref{tab:full cft} setting $N=0$, $p=0,1,2$, in the $\C^3/\Z_4$ orbifold.  Since nodes 0 and 2 are both empty, there are two exotic (one-)instanton contributions that add up to give the full non-perturbative superpotential.
The field content of the theory as well as the charged modes of the D(-1) are displayed in Figure \ref{fig:exoticinstanton} for the instanton contribution coming from node 0.
The couplings of matter fields with the instanton modes read:
\be 
S_{\rm charged} \sim \mu^i S \mu_i + \mu' {\cal M}  \mu',
\ee
where ${\cal M}$ is the expectation value of a non-dynamical D7-D7 field, transforming in the antisymmetric representation of the SU($2p$) (flavour) group at node 1. Notice that the non-dynamical fields ${\cal M}$ and $\tilde {\cal M}$ do not  produce any effect at the pertubative level because the nodes 0 and 2 are empty. Hence, vev's of these fields are perturbatively allowed without breaking conformal invariance.   For $p=0$ both $\mathcal M$ and $\mu'$ are absent.
The contribution to the effective action takes the schematic form:
\be
S_{\rm n.p.} \sim \int d^4x\ d^2\theta \int d^{6-2p}\mu\ d^{2p}\mu'\ e^{-S_{\rm charged}}.
\ee
There is only one way to saturate all fermion modes, which is to bring down a term $\mathcal M^p\,S^{3-p}$.
Taking into account the analogous one-instanton correction arising from node 2, one finds:
\be\label{exotic superpot U2}
W_{\rm exotic} \sim {\cal M}^p S^{3-p}+\tilde {\cal M}^p\tilde S^{3-p}.
\ee
For $p=0$ formula (\ref{exotic superpot U2}) produces Yukawa couplings  preserving conformal invariance. For $p=1,2$,
conformal invariance is dynamically broken at the scales set by ${\cal M}$ and $\tilde {\cal M}$. For $p=2$ a Polony-like
term is generated inducing supersymmetry breaking, too.  It's important to note that the  absence of a $\Lambda$ mass scale 
in (\ref{exotic superpot U2}) reflects the vanishing of the putative one-loop beta function coefficients of the two empty nodes: $\beta_0=\beta_2=0$.

As a second example, we consider a  conformal gauge theory in a $\Z_3$ quiver with an empty non-conformal node. 
 The model is displayed in the last row of Table \ref{tab:cft with empty node}. It admits an exotic instanton contribution arising from the empty `Sp(0)' node.
The matter content and D(-1) modes are again depicted in Figure \ref{fig:exoticinstanton} and the couplings with charged modes are as follows (we separate $\mu^I = (\mu^i,\ \tilde\mu)$ with $i=1,2$; $A_i,\ S$ sit in the antisymmetric and symmetric representations respectively):
\be
S_{\rm charged} \sim \mu^i\mu_i S + \tilde\mu \mu^i A_i + \tilde\mu\mu' Q + \mu'\mu' \mathcal M.
\ee
Similarly to the previous example, the mass scale ${\cal M}$ is the expectation value of the D7-D7 field  transforming in an antisymmetric representation of the SU(3) flavour group associated with D7 branes in node 1.
When $N$ is even or $N=1$ there is no contribution to the superpotential.
For odd $N\ge3$ one finds that there are two ways to saturate all fermion zero-modes, leading to
\be\label{exotic superpot Z3}
W_{\rm exotic} \sim \Lambda_0^{\frac32(1-N)}\left(Q^3 A^{N-3}S^{(N+3)/2}+ \mathcal M\, Q\, A^{N-1}S^{(N+1)/2} \right).
\ee
We notice that, unlike in the previous example, the exotic superpotential that breaks conformal symmetry is generated even when the vev of the D7-D7 field ${\cal M}$ is set to zero.
The presence of an overall scale $\Lambda_0$ in (\ref{exotic superpot Z3}), responsible for the breaking of conformal symmetry, reflects the fact that in this example the putative one-loop beta function of the empty node is non-zero: $\beta_0=\frac32(1-N)$.

Another interesting possibility is to have both gauge and exotic instanton contributions.
Looking at Table \ref{thz345inst}, we can see for instance that in the $\Z_4$, $\epsilon_0=+1$ models it's possible to set $N_2=0$ and obtain theories that exhibit one-instanton superpotential contributions both from a gauge instanton in the ${\rm Sp}(N_0)$ node and an exotic instanton at the Sp(0) node.

\subsection{Scales and closed string moduli}

As remarked above the scales $\Lambda$'s entering the superpotentials carry an explicit 
dependence on the closed string moduli $T_a$ describing the complex K\"ahler deformations of the singularity.
Their imaginary parts parametrize Fayet--Illiopolous terms for the gauge theory at the corresponding node of the quiver. 
 (Twisted) complex structure moduli $U_\alpha$, if present, are associated to 3-form fluxes
 and generate mass deformations of the quiver gauge theory\footnote{We are currently analysing this issue \cite{wipwithImp}}. 
The explicit form of the tree-level (disk) gauge kinetic functions $f_a(T_h)$ and thus of the RG invariant scales $\Lambda_a$ depends on the node where the `fractional' brane sits
\be
\Lambda_a=M e^{2\pi f_a(T_b) },
\ee
where $M$ is some (holomorphic) mass-scale.  
The fields $\mathrm{Im} T_b$  transform under U$(1)_a\subset $ U$(N_a)$ according to
\be
\delta_a \mathrm{Im} T_b = N_a ( w^{a b}  - w^{(n-a)b}) \alpha_a\quad \rm (no \ sum) ,
\label{axionic}
\ee
when
\be
\delta_a A_\mu^b = \delta_a^b\ \partial_\mu \alpha_a\quad \rm (no \ sum).
\ee
The axionic shifts (\ref{axionic}) compensate for the transformation properties of the chiral fields entering in the superpotential.   \ie the shift symmetry of the RR-axion  ${\rm Im}T_a$ is gauged by the `anomalous' U$(1)$ vector boson $A_\mu^b$.
As a result of the linear dependence of $f_a$ on $T_b$ 
\be
f_a  = \sum_{b,c=0}^{n-1} I_{a,b}\, w^{b c}\, T_c \,,
\ee
the gauging of the axionic shifts induces the following transformations of the holomorphic gauge kinetic functions 
\be
\delta_a f_b = N_a (I_{a,b}-I_{n-a,b}) \alpha_a\,.
\label{df}
\ee
For the first few $n$ one finds
\bea\label{ex}
n=3: &&   \delta_1 f_1=3 N_1\alpha_1,\nn\\[.6em]
n=4: &&   \delta_1 f_1= -  N_1 \alpha_1  
,\\[.2em]
\nn
n=5: &&   \delta_a f_b=
\left(
\begin{array}{cc}
N_1 \alpha_1 &  N_1 \alpha_1   \\
-3 N_2 \alpha_2 &   2 N_2  \alpha_2 \\
\end{array}
\right) \label{ffs}
\eea
and so on.

\section{S-dual quiver gauge theories }
\label{sdual}

In a recent paper \cite{GarciaEtxebarria:2012qx}, a new duality relating ${\cal N}=1$ unoriented quiver theories that is based on S-duality of the parent ${\cal N}=4$ unoriented theory has been proposed.
 
Indeed S-duality of type IIB theory can be used to relate the dynamics of different unoriented projections of (quiver) gauge theories living on D3-branes. In flat space-time
the U$(N)$ ${\cal N}=4$ SYM governing the low-energy dynamics of a stack of D3-branes is self-dual. The same is true for the SO$(2N)$ ${\cal N}=4$ SYM governing the low-energy dynamics of a stack of D3-branes on top of a `standard' $\Omega 3^-$ plane. If one however consider `exotic' $\Omega 3$-planes carrying non trivial (but quantized \cite{Bianchi:1991eu, Bianchi:1997rf, Witten:1997bs}) 2-form fluxes\footnote{Recall $\Pi_2(S^5/\Z_2) = \Z_2$ \cite{Witten:1998xy}.} the situation changes. $\Omega 3^+$ carrying $(B_2,C_2)=(1/2,0)$ and giving rise to Sp$(2N)$ is conjectured to be S-dual to $\Omega 3^-$ carrying $(B_2,C_2)=(0,1/2)$ and giving rise to SO$(2N+1)$. Finally $\Omega 3^+$ carrying $(B_2,C_2)=(1/2,1/2)$ and giving rise to Sp$(2N)$ is self-dual \cite{Witten:1998xy}. The last two are usually referred to as $\tilde\Omega 3^{\pm}$.

In  \cite{GarciaEtxebarria:2012qx}  the duality between  SO and Sp orientifolds have been extended to 
${\cal N}=1$ settings including the $\mathbb C^3/\mathbb Z_3$ unoriented quiver as well as non-orbifold toric singularities.
The duality proposal has been substantiated by a precise matching not only of the gauge-invariant degrees of  freedom and the anomalies of global symmetries but also of  dynamical effects taking place on the two sides of the duality.  Here we extend the analysis to the whole $\mathbb C^3/\mathbb Z_n$ series and propose an infinite sequence of new SO/Sp dual pairs of  unoriented quiver gauge theories without flavour.
We support the duality by matching the spectra of gauge invariant operators on the two sides of the duality.  In particular, we show that $\Omega 3^+$-plane can be replaced by $\Omega 3^-$-plane plus certain number of fractional D3-branes determined by  a simple geometric relation.
We restrict ourselves to the case with no D7  branes (nor $\Omega$7-planes). Adding D7-branes would naively spoil the duality since D7-branes transform non-trivial under S-duality\footnote{It would be interesting to explore similar duality relations in
presence of S-duality invariant configurations of mutually non-local 7-branes.}.  

For concreteness we take  $\C^3/\Z_n$ with $n$ odd. We look for SO/Sp orientifold quiver dual pairs in presence of a single $\Omega 3$ plane,
i.e. $\epsilon_I=({-\,}{-\,}{-})$. We denote by ${\bf N}=\{ N_a \}$ the number of fractional branes in the
Sp gauge theory and by ${\bf \tilde N} =\{ \tilde N_a \}$ that in the SO gauge theory. Cancellation of anomalies in the two gauge theories requires
\bea
I\cdot {\bf N}+4\,K=I\cdot {\bf \tilde N}-4\, K=0.
\label{anomnof}
\eea 
The two equations are solved by
\bea
{\bf N} &=&c_+ \, {\bf v} -4\, I_\perp^{-1} \cdot K\nn\\
{\bf \tilde N} &=&  c_- \, {\bf v}+  4\, I_\perp^{-1} \cdot K,
\label{solNn}
\eea
with  ${\bf v}=(1,1,...)$ and $c_\pm$ arbitrary.  By $I_\perp^{-1}$ we denote the inverse of $I$ in the space orthogonal to ${\bf v}$. 
We notice that terms proportional to ${\bf v}$ in (\ref{solNn}) do not contribute to (\ref{anomnof})  since $I\cdot {\bf v}=0$,  or in other words anomaly equations are not modified by the addition of regular branes.  To fix $c_\pm$ we recall that before the orbifolding $\Omega 3^+= \Omega 3^-+1\, D3$ and so the total number of fractional branes in the SO gauge theory should exceed by one that  in the Sp theory, i.e.  $ {\bf v} \cdot ({\bf \tilde N}-{\bf N})=1$, which translates into $c_- - c_+=\ft{1}{n}$.
In addition, one should require that ${\bf N}$ and ${\bf \tilde N}$ are made of integers. 
The solution is parametrized by an integer $p$ and can be written as
\be
 c_{\pm} =p+\ft{1}{2} \mp \ft{1}{2n}. 
\ee
One can easily check that ${\bf N}$ and ${\bf \tilde N}$ given by (\ref{solNn}) are always integers and positive for $p$ large enough. The resulting gauge theory for $n=3,4,5$ are displayed in table \ref{tablesospD3}.
The case $n=3$ reproduces the series of dual pairs studied in \cite{GarciaEtxebarria:2012qx}.

\begin{table}[h!]
\centering
\begin{tabular}{cll}
\toprule
  & Gauge theories  & d.o.f.  
\\ \otoprule
\multirow{2}{*}{$\Z_3$}  
   &  ${\rm Sp}(2p+4)\times {\rm U}(2p)$ 
   & $\nu_I = 9p + 6p^2$, 
\\
   &   ${\rm SO}(2p-1)\times {\rm U}(2p+3)$ 
   & $\nu_0 =10+ 9p + 6p^2$
\\ \midrule
\multirow{2}{*}{$\Z_5$}  
   &  ${\rm Sp}(2p+2)\times {\rm U}(2p+2)\times {\rm U}(2p-2)$ 
   & $\nu_{1,2} = 1+ 5p + 10p^2$,
\\   
   &  ${\rm SO}(2p-1)\times {\rm U}(2p-1)\times {\rm U}(2p+3)$
   & $\nu_{3} = \nu_{1,2}-6,\ \nu_0 = \nu_{1,2}+10$
\\ \midrule
\multirow{2}{*}{$\Z_7$} 
   & ${\rm Sp}(2p+8)\times {\rm U}(2p+4)^2\times {\rm U}(2p)$ 
   & $\nu_{1,2} = 48+49p + 14p^2$,
\\
   &  ${\rm SO}(2p-1)\times {\rm U}(2p+3)^2\times {\rm U}(2p+7)$
   & $\nu_{3} = \nu_{1,2}-6,\ \nu_0 = \nu_{1,2}+20$
\\ \bottomrule
\end{tabular}
\caption{Examples of Sp/SO dual models, with $N_a\ge1$.}
\label{tablesospD3}
\end{table}

In the rest of this section we collect  some evidences for the duality between SO/Sp quiver gauge theories  with fractional brane content (\ref{solNn}) on a general $\C^3/\Z_n$ orientifold singularity.
The main check relies on the comparison of the spectra of the  two gauge theories. 
To this aim, we organize the states of the two gauge theories according to their charges with respect to the global U$(1)^3$ symmetries. U$(1)^3$ is  the Cartan of the SO$(6)$ R-symmetry of the parent ${\cal N}=4$ theory and it is therefore part of the global symmetry of any orientifold theory.
There are three types of chiral multiplets ${\bf C}^I$, each one charged  respect to one U$(1)_I\in$ U$(1)^3$. We denote by $\nu_I$ the number of degrees of freedom of each.
Gauge invariant degrees of freedom are built out of traces involving these fields.
This leads to  $\sum_I \nu_I-{\rm dim} G$  mesonic/baryonic degrees of freedom.
In the Sp gauge theory one finds
\bea
  C^I :   &&      \nu_I= \sum_a \left(  N_a N_{a+a_I} +\epsilon  N_a \delta_{a+a_I,-a}\right),  \nn\\
  {\rm dim} G: &&  \nu_0=-\left( \ft12 \sum_a N_a^2 +\ft12 \epsilon \, N_0 \right),
\eea
with $\epsilon=+$. The spectrum of the SO gauge theory on the other hand is given by the same formulas with $\epsilon=-$ and $N_a\to \tilde N_a$. The difference of degrees of freedom between the two gauge theories is
\bea
 \Delta \nu_I &=& \sum_a  (N_a+\tilde N_a) \left(    \tilde N_{a+a_I}-  N_{a+a_I}   - \delta_{a+a_I,-a}\right) =0, \nn\\
 \Delta \nu_0 &=&- \ft12 \sum_a (\tilde N_a+ N_a) (\tilde N_a- N_a) +\ft12 \, (N_0+\tilde N_0)=0 , \label{deltas}
\eea
where in the right hand side we used  (\ref{solNn}),  to write $\tilde N_a+ N_a=(2p+1) {\bf v}$, $({\bf \tilde N}-{\bf N})\cdot {\bf v}=1$. We notice that the matching between the degrees of freedom $\nu_I$ automatically ensures the matching of anomalies involving the U$(1)^3$ symmetries and therefore is a  strong support of the claimed duality relation between the SO and Sp gauge theories.
In the last column of Table \ref{tablesospD3} we display the number of degree of freedom for the first few candidates of dual pairs.

We remark that the relation between the SO and Sp gauge theories can be translated into a purely geometric identification between the  cycles wrapped by  the ${{\Omega 3}}^+$ and ${\Omega 3}^-$ planes in the two theories. Indeed, it corresponds to the identification   
\be
{\Omega 3}^+={\Omega 3}^- + (\tilde{N}_a-N_a) D3_a   ,
\label{opom}
\ee  
with $(\tilde{N}_a-N_a)$ such that the cycles wrapped by the two ${\Omega 3}$-planes coincide:
\be
4\, \pi_{\Omega 3}=-4\, \pi_{\Omega 3} + (\tilde{N}_a-{N}_a) \, \pi_{D3a}   .
\ee
Multiplying (\ref{opom}) and using (\ref{prodos}) one finds agreement with 
(\ref{anomnof}). In addition, one requires that $\sum_a (\tilde{N}_a- N_a)=1$ to match the duality in the parent theory in flat spacetime. 

Although the matching of the dof's, including their non-anomalous flavour charges, seems to be only a necessary condition, we believe this is equivalent to matching all triangle anomalies as carefully done in \cite{GarciaEtxebarria:2012qx} for the $\Z_3$ case. 
Indeed, matching of $\nu_I$ and $\nu_0$ implies that the number of  gauge invariant operators matches on the two sides of the duality. In particular chiral operators that contribute to the superconformal index should match \cite{Romelsberger:2005eg, Romelsberger:2007ec, Dolan:2008qi}.   
We observe that the matching of gauge invariant degrees of freedom can be traced in the analogous matching before the orbifold projection. If we denote by $\Phi^I$ and $\tilde \Phi^I$ the 3 chiral multiplets in the adjoint of Sp$(2N)$ and SO$(2N+1)$ (before the orbifold projection) one can see that the number of singlets one can built at each dimension in the two gauge theories matches perfectly. This implies
the correspondence
\be
{\rm Tr} (\Phi^{I_1} \Phi^{I_2} \ldots \Phi^{I_k} ) \leftrightarrow  {\rm Tr} (\tilde\Phi^{I_1} \tilde\Phi^{I_2} \ldots \tilde\Phi^{I_k} ) \label{trphi}.
\ee
In the quiver gauge theory, gauge invariant operators are given again by (\ref{trphi})  with  $\Phi^I$ and $\tilde \Phi^I$ now given by block matrices satisfying the orbifold invariant conditions.
Moreover the basis of gauge invariant operators has $\sum_I \nu_I-\nu_0$ elements for the two dual gauge theories.
Further dynamical checks of duality, including but not limited to a detailed comparisons of the superconformal indices, may help identifying the class of unoriented quiver dual pairs.

\section{Conclusions}

We have discussed unoriented quiver theories with flavour that govern the low-energy dynamics of D3-branes at orbifold singularities in the presence of (exotic) $\Omega$ planes and D7-branes wrapping non-compact cycles. The presence of a net number of `fractional' branes, as compared to the case without  $\Omega$ planes and D7-branes, makes the theories intrinsically chiral, generically non superconformal and thus phenomenologically more promising than theories with only `regular' D3-branes. 

In the recent past oriented quivers for D3-branes at toric singularities that admit a dimer description have received a lot of attention. Although orientifolds of dimers have already been analysed in \cite{Franco:2007ii}, here we have tackled  the problem from a world-sheet perspective in the restricted context of non-compact $\Z_n$ orientifolds.   

We have rederived the relation between tadpole and anomalies, taking into account the flavour branes, identified the locally consistent embeddings of the D7 and the various allowed unoriented projections. 

We have then recognized the conditions for restoring superconformal invariance in the presence of D7 and $\Omega$ planes, focusing on two classes of models with $\beta_a =0$ at all nodes and with $\beta_a =0$ at all but one `empty' node. We have relied on previous analyses of `dilaton' tadpoles and RG flows, in order to argue that no anomalous dimensions are expected for the matter fields that would require consideration of the NSVZ `exact' $\beta$ function rather than our simple-minded one-loop $\beta$ function.

We have also classified quiver theories that receive non-perturbative corrections to the superpotential from unoriented D-brane instantons of the `gauge' or `exotic' kinds. In particular we have found a theory where both kinds of corrections are present and conformal theories where the conformal symmetry is broken dynamically via the generation of exotic superpotentials.

We have finally turned our attention on to the recently proposed ${\cal N} =1 $ duality, which is a remnant of the ${\cal N}=4$ S-duality between Sp$(2N)$ and SO$(2N+1)$ gauge groups. We have identified candidate dual pairs and given further evidence for the validity of the duality in the orbifold context. 

It would be interesting to study the effect of 3-form NSNS and RR fluxes on the gauge theory dynamics. In particular, this can result in moduli stabilisation and topology changes. Indeed one can show that some orbifold singularities with vector-like matter can be connected to more general non-orbifold singularities. Work on this issue is in progress \cite{wipwithImp}. 

Another issue is related to the global embedding of the unoriented quivers with flavour. The consistent gauging of the D7-brane flavour symmetry requires the absence of chiral anomaly, and thus of global tadpoles, as well as other subtler, K-theoretic, issues that have been recently addressed for instance in \cite{Cicoli:2013mpa} for the case of two oriented quiver theories with flavour on $\Z_3$ singularities exchanged by an orientifold projection.

\vspace{1cm}

\centerline{\large\bf Acknowledgments}

We would like to acknowledge A.~Amariti, C.~Bachas, S.~Cremonesi, E.~Dudas, S.~Franco, F.~Fucito, A.~Hanany, M.~Petropoulos, G.~Pradisi, R-K.~Seong, Y.~Stanev, G.~Travaglini  for interesting discussions and above all L.~Martucci for collaboration at an early stage of this project.   
 This work is partially supported by the ERC Advanced Grant n.226455 Superfields and by the Italian MIUR-PRIN contract 2009-KHZKRX.
The work of D.~R.~P. is also supported by the Padova University Project CPDA119349.   
G.~I. thanks the University of Amsterdam for hospitality during the early stages of this project.  
While this work was being carried on, M.~B. has been visiting Imperial College (IC) in London, Queen Mary University of London (QMUL), Ecole Normale Superieure (ENS) Paris and Ecole Polytechinque (EPoly) Paris. M.~B. would like to thank A.~Hanany, G.~Travaglini, C.~Bachas, M.~Petropoulos, J.~Iliopoulos and their colleagues for their very kind hospitality, for creating a stimulating environment and to acknowledge partial support trough Internal EPSRC Funding (IC), Leverhulme Visiting Professorship (QMUL), Visiting Professorship (ENS, EPoly) and Institut Philippe Meier (ENS). \vspace{0.5cm}

\begin{appendix}

\section{ String partition function}

In this appendix we review the computation of the string partition function for a system of unoriented closed and 
open strings on $\C^3/\Z_{n} $. See \cite{Angelantonj:2002ct} and references therein for a general review on open and unoriented strings .

\subsection{Torus amplitude}
\label{torusappendix}
 
Closed string states organize into $g$-twisted sectors defined by
the boundary conditions     
\be
X^I (\sigma+2\pi ,\tau)=w^{g a_I} X^I (\sigma,\tau)   ,
\ee 
with similar conditions for fermions. The torus partition function can be written as $\int {d^2 \tau_2\over \tau_2^2}$  times\footnote{For simplicity we normalize to one the infinite 
volumes of each complex plane.}
\be
\label{torus}
{\cal T}=\sum_{g,h=0}^{n-1}  {\cal T}_{g,h}={1\over n} \sum_{g,h=0}^{n-1} |\rho[^g_h] (\tau)|^2 \Lambda[^g_h] (\tau,\bar \tau),
\ee
with
\be
\Lambda[^g_h]= 
\int d p\, e^{-\pi \tau_2 p^2} \langle p| \Theta^h |p\rangle
\ee
the contribution of zero modes  momenta (along the plane invariant under $\Theta^g$) and
\be
\label{rhodef}
\rho[^g_h](\tau) ={\rm Tr}_{\rm g-twisted} \left[ \left({1+(-)^F\over 2}\right)   \Theta^h \, q^{L_0-a}\,\right]
\ \ \ \ \ \ \ \ q=e^{2\pi i\tau}
\ee
the oscillator part of the $h$-projected chiral partition function in the $g$-twisted sector.  
Explicitly
\bea
\label{rhodef2}
\rho[^g_h](\tau) &=& \ft12\sum_{a,b=0}^1 (-)^{ a +b+ ab}   \prod_{I=0}^3 
  {\vartheta\left[^{a+{2g a_I\over n}  }_{b+{2h a_I \over n}  }\right]\over 
\vartheta\left[^{1+{2g a_I\over n} }_{1+{2h a_I\over n} }\right] }    \,
    \prod_{I\in {\cal C}_{g,h}}2 \sin(\ft{\pi h a_I}{n})  \nn\\
&=&
-\left(  {\vartheta_1\over \eta^3} \right)^{\cal N}       \,   \prod_{I\in {\cal C}_{g,h}}2\sin(\ft{\pi h a_I}{n}).
\eea
The product in the second line  runs over all $I$'s  such that $ g a_I \in  n \Z$ but  $ h a_I \notin  n\Z$.
We denote this set by ${ \cal C}_{g,h}$.
The   terms $2\sin(\pi h a_I)$ cancel the zero mode part of the  corresponding theta functions in the denominator. ${\cal N}$ counts the number of complex planes invariant under both $\Theta^g$ and $\Theta^h$, i.e. those planes I satisfying $ g a_I, h a_I \in n \Z $. We notice that ${\cal N} $ is also the number of supersymmetries preserved by the $\Theta^g,\Theta^h$ twists.
In the second line of (\ref{rhodef2}) we used the Jacobi identity to perform the spin structure sum. Notice that only fermionic zero modes contribute to (\ref{rhodef2}), since $\vartheta_1 \approx (1-1) \eta^3$.
Here and below indices $a,b$ are understood modulo $n$, e.g. $\delta_{a,0}$ means $a\in n \Z$, and so on.\\
On the other hand the bosonic zero mode contribution is given by 
\bea
\Lambda[^g_h] &=& 
{1\over \tau_2^{\cal N} } \prod_{I \in{ \cal C}_{g,h} } \int d^2 p_I \, e^{-\pi \tau_2 p_I^2} \langle p_I| \Theta^h |p_I\rangle=
{1\over \tau_2^{\cal N} } \prod_{I \in{ \cal C}_{g,h} } \int d^2 p_I \delta( p_I-w^{h a_I} p_I)\nn\\
&=& {1\over \tau_2^{\cal N} } \prod_{I \in{ \cal C}_{g,h} }    \, {1\over  |1-w^{h a_I} |^{2   }  }.
\label{lamb1}
\eea
Plugging (\ref{rhodef2}) and (\ref{lamb1}) into (\ref{torus}) one finds that bosonic and fermionic
zero mode contributions cancel against each other and one is left with the result
\be
{\cal T}_{g,h}={1 \over \tau_2^{\cal N} }     \left|
{\vartheta_1 \over \eta^{3}}  \right|^{2{\cal N}}  .
\label{tgh}
\ee
Remarkably, the result (\ref{tgh}) depends only on the number ${\cal N}$ of supersymmetries preserved by the twists and not on the details of $g$, $h$.
The torus amplitude can then be written in the simple form
\be
{\cal T}={1\over n} \sum_{{\cal N}=1,2,4} \sum_{[^g_h]\in {\rm Orb}_{\cal N} } {1 \over \tau_2^{{\cal N} }}   \label{torus2}
\, \left|
{\vartheta_1 \over \eta^{3}}  \right|^{2{\cal N}},
\ee
where we denote by ${\rm Orb}_{\cal N}$ the set of twists $g,h$ leaving invariant ${\cal N}$ out of the four complex coordinates. 
We are interested in the physics localized around the singularity. States localized around the singularity come from  ${\cal N}=1$ sectors where all three coordinates $X^I$ are twisted, i.e. $g a_I\notin \Z$ for any $I$.   States in ${\cal N}=2,4$ sectors are non normalizable and will be discarded in the following. 

We notice also that the result (\ref{torus2}) is modular invariant since $\vartheta_1 \over \eta^{3}$ is invariant under $T$ and transforms as 
\be
S: \qquad  {\vartheta_1 \over \eta^{3}} \to {1\over \tau } {\vartheta_1 \over \eta^{3}}
\ee
under the S-modular transformation. 

\subsection*{Helicity traces}

In this paper we deal with partition functions of supersymmetric theories that are always zero due to the matching between
the number of bosonic and fermionic degrees of freedom in these theories. In particular, the partition function in sectors with ${\cal N}$ supersymmetries vanishes as $(1-1)^{\cal N}$, indicating that multiplets in these theories contain $2^{{\cal N}-1}$
bosons and  $2^{{\cal N}-1}$ fermionic states. It is often convenient to resolve this degeneracy by counting states  weighted
by their helicity on a plane.  In this way one can distinguish between vector and chiral multiplets in ${\cal N}=1$ or
vector and hypermultiplets in ${\cal N}=2$ theories. To define the helicity trace of string states one can simply
replace the chiral partition functions $\rho_{gh}$ by the character value function
\be
\rho_{gh}(x)=  2\sin \pi x  {\vartheta_1(\ft{x}{2})\over \vartheta_1(x)}
\prod_{I=1}^3   \left(  \left(2\sin \ft{ \pi h a_I}{n} \right)^{\delta_{g a_I,0}}    {\vartheta\left[^{1+{2g a_I\over n} }_{1+{2h a_I\over n} }\right] (\ft{x}{2})
\over  \vartheta\left[^{1+{2g a_I\over n} }_{1+{2h a_I\over n}} \right](0)  }  \, \right ).
\ee
In particular, for $g=h=0$ one finds 
\be
  {\cal N}=4:\qquad\rho[^0_0]\sim \sin^4\left( \ft{\pi x}{2}\right)+O(q)=e^{2\pi  {\rm i} x}+e^{-2\pi  {\rm i} x}+6-4( e^{\pi  {\rm i} x}+e^{-\pi  {\rm i} x}  )+O(q),
\ee
reproducing the helicity content of an ${\cal N}=4$ multiplet. Similarly, for  $[^g_0] \in {\rm Orb}_{\cal N}$ with ${\cal N}=1,2$ one finds
\bea
{\cal N}=2:\qquad   \rho[^g_0] &\sim& \sin^2\left( \ft{\pi x}{2}\right)+O(q)\sim2-e^{\pi  {\rm i} x}-e^{-\pi  {\rm i} x} +O(q)\nn,\\
 {\cal N}=1:\qquad    \rho[^g_0]  &\sim& \sin\left( \ft{\pi x}{2}\right)+O(q) \sim 1-e^{\pi  {\rm i} x}  +O(q),
\eea
reproducing the helicity content of the ${\cal N}=2$ hyper and ${\cal N}=1$ chiral multiplets respectively.
We will mainly focus on ${\cal N}=1$ sectors proportional to $\vartheta_1(\ft{x}{2}) \sim  \sin\left( \ft{\pi x}{2}\right)$. The coefficient of $\vartheta_1$ should be interpreted as the net number of chiral fields, i.e. as the difference between spinors of left and right moving chirality in the open string spectrum. 

\subsection{Klein bottle amplitudes}

We now consider the inclusion of  an $\Omega$-plane at the singularity.
This corresponds to quotienting the type IIB string theory at the singularity by an action $\Omega_\epsilon$ involving a worldsheet parity and a reflection specified  by four signs $(\epsilon_0,\epsilon_I)$ satisfying $\prod_{I=1}^3 \epsilon_I=-1$. 
 
The Klein bottle amplitude is given by the insertion of  $\Omega_\epsilon$ in the torus amplitude. It is important to notice that only $\Omega$-unpaired states can contribute to the Klein bottle amplitude.
In particular, $g$-twisted sectors combine left moving states with their complex conjugate and therefore can contribute to the Klein only if they come in real representations i.e. either for $g=0$ or $g={n \over 2}$ in the case of even $n$. 
Inserting $\Omega_{\epsilon}$ in the momentum integral (\ref{lamb1}) one finds
\bea
\Lambda^{\Omega}[^g_h] &=& 
{1\over \tau_2^{\cal N} } \prod_{I \in{ \cal C}_{g,h} } \int d^{2} p_I e^{-\pi \tau_2 p_I^2} \langle p_I | \epsilon_I w^{a_I h} p_I \rangle=
{1\over \tau_2^{\cal N} } \prod_{I_{g,h} }{1 \over |1- \epsilon_I w^{a_I h} |^2 }   .
\label{lamb2}
\eea
Combined with contributions coming from the diagonal part $\rho[^g_h](2 i t)$ one finds 
\bea
{\cal K}_{0,h} &=&  - \,   \prod_{I=1}^3 {( 1-  w^{2a_I h} ) \over   (1- \epsilon_I w^{a_I h} )^2 }   \int {dt\over t^3}  \,{\vartheta_1\over \eta^3}(2 {\rm i}t) , \nn\\
{\cal K}_{{n\over 2},h} &=&  -\hspace*{-.5em}\prod_{I:\, a_I\, \rm even} \frac{(1- w^{2a_Ih})}{(1-\epsilon_I w^{a_Ih})^2}
\int {dt\over t^3}  \,{\vartheta_1\over \eta^3}(2 {\rm i}t),
\eea
for the $\Theta^h$-projected amplitudes in the $g=0$ and $g=\ft{n}{2}$ twisted sectors respectively and zero otherwise.

\subsection{Annulus and Moebius strip amplitudes}

Finally we consider  the inclusion of fractional D3 and D7-branes  at the singularity. Fractional branes are classified by the representations ${\bf R}_a$ of $\Z_n$ with $a=0,...n-1$. We denoted by $N_a(M_a)$ the number of fractional D3(D7) branes of each type and by N(M) the total number. We are interested in the low energy of the four-dimensional theory localized at the singularity described by  open string states with at least one end on D3 branes. The dynamics of these states is described by an effective ${\cal N}=1$ supersymmetric quiver gauge theory.  We orient D7 branes along the $I=1,2$ planes.

\subsubsection*{The action on Chan-Paton indices}
The full action of the orbifold and orientifold projections gives the following identifications for the Chan-Paton matrices $\lambda$ associated to D3-D3 and D3-D7 fields:
\begin{align}
\label{explicit orbifold action}
\Theta^h:\ \ 
& \lambda_{\mathbf V} = \gamma_{\Theta,\mathrm D3}  \lambda_{\mathbf V} \gamma_{\Theta,\mathrm D3}^{-1}
&  \lambda_{\mathbf C^I} &= w^{a_I}\gamma_{\Theta,\mathrm D3} \lambda_{\mathbf C^I} \gamma_{\Theta,\mathrm D3}^{-1}
& \lambda_{ \mathbf C_{3,7}^{\dot a}} &= w^{s}\gamma_{\Theta,\mathrm D3} \lambda_{\mathbf C_{3,7}^{\dot a} }\gamma_{\Theta,\mathrm D7}^{-1}\\[.5em]
\label{explicit Omega action}
\Omega\ :\ \ 
&\lambda_{ \mathbf V} = - \gamma_{\Omega,\mathrm D3} \lambda_{\mathbf V}^T \gamma_{\Omega,\mathrm D3}^{-1}
& \lambda_{\mathbf C^I} &= \epsilon_I \gamma_{\Omega,\mathrm D3} (\lambda_{\mathbf C^I})^T \gamma_{\Omega,\mathrm D3}^{-1}
& \lambda_{\mathbf C_{3,7}^{\dot a}} &=  \gamma_{\Omega,\mathrm D3} (\lambda_{\mathbf C_{7,3}^{\dot a}})^T \gamma_{\Omega,\mathrm D7}^{-1}.
\end{align}
Up to some choices of phases and conventions, one can write the explicit embedding of the projections in the Chan--Paton group:
\bea
\gamma_{\Theta,\rm D3} &=&
\left(
\begin{array}{ccccc}
\mathbf{1}_{N_0} &&&&\\
&w\, \mathbf{1}_{N_1}&&&\\
&&w^2\, \mathbf{1}_{N_2}&&\\
&&&&\hspace*{-1.5em}\ddots\\
\end{array}
\right)\,,  
\label{gampro1}  
\\[.5em]
\gamma_{\Omega,\rm D3}  &=&
\left(
\begin{array}{cccccc}
\Delta_{N_0} & &&&&\\
&0&\cdots&\cdots&0&c\mathbf1_{N_1}\\
&\vdots&&&c\mathbf1_{N_2}&0\\
&\vdots&&\iddots&&\vdots\\
&0&c^*\mathbf1_{N_{2}}&&&\vdots\\
&c^*\mathbf1_{N_{1}}&0&\cdots&\cdots&0\\
\end{array}
\right)\,,
 \qquad
\begin{array}{rcl}
\Delta_{N_a}&=&
\left\{
\begin{array}{lc}
\mathbf{1}_{N_a}&\epsilon_0=-1\\
 {\rm i} J_{N_a}&\epsilon_0=+1\\
\end{array}\right.\\
\vphantom{\prod}&& \nn\\
c&=&\left\{
\begin{array}{lc}
1&\epsilon_0=-1\\
 \,{\rm i}&\epsilon_0=+1\\
\end{array}\right.
\end{array}
\eea
where $J_{N_a}$ is the (real, antisymmetric) quadratic invariant  of Sp$(N_a)$.
When $n$ is even, the central entry of the antidiagonal  block in $\gamma_{\Omega,\rm D3}$ corresponds to the second SO/Sp node of the quiver and therefore is of the form $\Delta_{N_{n/2}}$.
In the case of $n$ even and $n/2$ odd, there is another inequivalent projection, corresponding to the identification of the node $0$ with the node $n/2$. 
The first example of this kind is $\Z_6$, where we can write:
\be
\gamma_{\Omega,\rm D3}=
\left(
\begin{array}{cccccc}
0 &\cdots & 0&c\mathbf1_{N_0} & & \\
\vdots & & c\mathbf1_{N_1} &0& & \\
 0& c^*\mathbf1_{N_1} & & \vdots & & \\
  c^*\mathbf1_{N_0} &0 &\cdots &0 & & \\
  &  & & & 0& c\mathbf1_{N_5}  \\
  &  & & &  c^*\mathbf1_{N_5} & 0
\end{array}
\right)
\label{gampro2}
\ee
These matrices satisfy the consistency condition 
\be\label{Omega D3 consistency}
\gamma^T_{\Omega,\mathrm D3} = -\epsilon_0 \gamma_{\Omega,\mathrm D3},
\ee
which can be obtained applying \eqref{explicit Omega action} twice.
This choice of sign combined with \eqref{explicit Omega action} provides the correct gauge group and matter field projections in the D3-D3 sector.
Consistency also requires that the D7-D7 sector exhibits the opposite unoriented projection:
\be\label{Omega D7 consistency}
\gamma^T_{\Omega,\mathrm D7} = \epsilon_0 \gamma_{\Omega,\mathrm D7},
\ee
which means that the same expressions (\ref{gampro1}) can be used for $\gamma_{\Theta,\rm D7}$, $\gamma_{\Omega,\rm D7}$ after replacing $N_a\to M_a$ and $\epsilon_0 \to -\epsilon_0$.
In the following we will use the shorter notation $\gamma_{h}\equiv\gamma_{\Theta,\mathrm{D}p}^h$ and $\gamma_{\Omega h}\equiv\gamma_{\Omega,\mathrm Dp}\, \gamma_{\Theta,\mathrm{D}p}^h$.

\subsubsection*{The annulus amplitude}

Let us first consider the annulus amplitude. There are three types of open strings depending on the boundary conditions at the two ends of the open string. Contributions from D3-D3 and D7-D7 open strings are proportional to the untwisted amplitude $\rho[^0_h]$. 
Finally D3-D7 open strings are twisted along the four-dimensional plane with mixed Neumann--Dirichlet boundary conditions and therefore have neither bosonic nor fermionic zero modes along this plane.
Collecting the various contributions one finds 
\bea
{\cal A}_h &=&  
- \prod_{I=1}^3 ( 1-  w^{a_I h} )  \left [  {\rm tr} _{\rm D3}\gamma_h -\frac{  w^{-{a_3 \over 2}h} {\rm tr} _{\rm D7}\gamma_h}{\prod_{I=1}^2 ( 1-w^{a_I h} )}\right]^2 \int {dt\over t^3} \, {\vartheta_1\over \eta^3} (\ft{ {\rm i}t}{2}) 
\label{directannulus}.
\eea
The three terms in the expansion of the square origin from D3-D3, D3-D7 and D7-D7 open strings respectively. $w$-dependent terms in the numerator and denominators come from contributions from fermionic and bosonic zero modes respectively.
Finally ${{\vartheta}_1\over { \eta}^3}  $ comes from bosonic and fermionic excitations transverse to the singularity.
The Chan Paton traces are
\be
{\rm tr} _{\rm D3}\gamma_h = \sum_{a=0}^{n-1} N_a w^{a h} ,\qquad 
{\rm tr} _{\rm D7}\gamma_h = \sum_{a=0}^{n-1} M_a w^{a h}  .\label{chan}
\ee

\subsubsection*{The Moebius amplitude}

The insertion of $\Omega_\epsilon$ in the D3-D3 and D7-D7 annulus leads to the Moebius amplitudes
\be
{\cal M}_h = \prod_{I=1}^3 ( 1+ \epsilon_I  w^{a_I h} )    \left[ 
{\rm tr}_{\rm D3} (\gamma^{-1}_{\Omega h}\gamma^{T}_{\Omega h})  
+\frac{  w^{-{a_3 }h} {\rm tr}_{\rm D7} (\gamma^{-1}_{\Omega h}\gamma^{T}_{\Omega h})  }{\prod_{I=1}^2 ( 1-w^{2 a_I h} )}  \right]  \int {dt\over t^3} 
\, { {\vartheta}_1\over { \eta}^3} (\ft{ {\rm i}t}{2}+\ft12) \label{directmoebius},
\ee
with
\be
{\rm tr}_{\rm D3} (\gamma^{-1}_{\Omega h}\gamma^{T}_{\Omega h})  =  -\epsilon_0 {\rm tr}_{\rm D3} \gamma_{2h},
\qquad
{\rm tr}_{\rm D7} (\gamma^{-1}_{\Omega h}\gamma^{T}_{\Omega h})  =  \epsilon_0 {\rm tr}_{\rm D7} \gamma_{2h}
\label{omegagamma}
\ee
for the unoriented projection defined by (\ref{gampro1}).

\subsubsection*{The spectrum of open string states}

The spectrum of the quiver gauge theory is codified in the Annulus and Moebius amplitudes.
States in vector multiplets come from open strings connecting D3 branes of the same type and realize the gauge symmetry  with orthogonal and symplectic gauge groups for nodes $a=0,\ft{n}{2}$ and unitary groups otherwise. 

Chiral multiplets come from open strings connecting D3 branes of different types or D3-D7 strings.
  They are summarized in the open string partition function
\bea
{1\over 2n} \sum_{h=0}^{n-1}{\cal A}_{h,D3D3+D3D7}& =&
-  \sum_{a,b=0}^{n-1} \left( \ft12 I_{ab} N_a  N_b  +J_{ab} N_a  M_b  \right)
\int {dt\over t^3} \, {  \vartheta_1\over \eta^3} (\ft{{\rm i}t}{2}), \nn\\
{1\over 2n} \sum_{h=0}^{n-1}{\cal M}_{h,D3}& =&
-\epsilon_0 \sum_{a=0}^{n-1}   \ft12  K_{a}  N_a   \int {dt\over t^3} \, {   \vartheta_1\over \eta^3} (\ft{{\rm i}t}{2}+\ft12)  ,
\label{spectrumam}
\eea
with
\bea
I_{ab} &=& \sum_{I=1}^3 ( \delta_{a,b-a_I} -\delta_{a,b+a_I}) ,\nn\\
J_{ab} &=&  \delta_{a,b-s} -\delta_{a,b+s},\nn\\
K_a &=& \sum_{I=1}^3  \epsilon_I ( \delta_{2a,a_I} -\delta_{2a,-a_I}  )
\eea
codifying the intersection numbers of the exceptional cycles at the singularity.
In deriving (\ref{spectrumam}) we repeatedly used the identity
\be
\prod_{I=1}^3 ( 1+\epsilon_I  w^{a_I h} )=\sum_{I=1}^3  \epsilon_I (w^{a_I h}  -w^{-a_I h} ).
\label{id}
\ee
 Using the fact that $\vartheta_1\sim 1-1$ counts the degrees of freedom of an ${\cal N}=1$ chiral multiplet (a vector
 multiplet is non-chiral) we conclude that the spectrum of chiral multiplet for the quiver can be written as
\bea
\label{orient2}
{\cal H}^{\rm open}_{\rm chiral} &=&  \sum_{a,b=0}^{n-1}\, \left(  \ft12 I_{ab} \fund_{a} \, \overline{\fund}_b +
   J_{ab}\, M_b \, \fund_{a}    \right) +   \epsilon_0  \sum_{a=0}^{n-1}\, \ft12 K_{a} \fund_{a}.
\eea

\subsection{Tadpole cancellation}

\subsubsection*{Odd $n$}  
The Klein, Annulus and Moebius amplitudes can be rewritten as cylinder amplitudes representing the exchange of a closed twisted string state between $\Omega$-planes and D-branes. We denote by $\tilde {\cal K}_{0,h}$, $\tilde {\cal A}_h$, $\tilde {\cal M}_h$ the corresponding amplitudes. The length $\ell$ of the cylinder is related to the one-loop modulus  $t$  via $\ell=({1\over 2t},{2\over t},{1\over 2t})$ for $({\cal K},{\cal A},{\cal M})$ respectively.  The ${\cal K}$ and ${\cal A}$ direct and transverse amplitudes are related by an $S$ modular transformation while the Moebius amplitudes are linked by $P=T S T^2 S$. Using
\begin{equation}
\begin{aligned}
S:  & \qquad {\vartheta_1\over \eta^3}\left( -{1\over \tau} \right) = {1\over \tau} {\vartheta_1\over \eta^3}(\tau),\\[.2em]
P:  & \qquad {\vartheta_1\over \eta^3}\left(   \frac{  {\rm i}}{2 \tau_2}+\frac{1}{2}  \right)   =  {1\over  {\rm i} \tau_2}  {\vartheta_1\over \eta^3} \left(   \frac{ {  {\rm i}}\,  \tau_2}{2 }+\frac{1}{2}   \right),
\end{aligned} 
\end{equation} 
one finds
\bea
\tilde{\cal K}_h &=&   {\rm i}\, 2^{2}   \,  \prod_{I=1}^3 {( 1-  w^{2a_I h} ) \over   (1- \epsilon_I w^{a_I h} )^2 } 
\int d\ell  \,{\vartheta_1\over \eta^3} (i \ell)  , \nn\\
\tilde{\cal A}_{h} &=&   {\rm i}\,  2^{-2}  \prod_{I=1}^3 ( 1-  w^{a_I h} )  \left [  {\rm tr} _{\rm D3}\gamma_h -\frac{  w^{-{a_3 \over 2}h} {\rm tr} _{\rm D7}\gamma_h}{\prod_{I=1}^2 ( 1-w^{a_I h} )}\right]^2 
\int d\ell  \, {\vartheta_1\over \eta^3} (i \ell)    \label{tadpol} ,\\
\tilde{\cal M}_h &=&   {\rm i} \, 2 \, \epsilon_0  \prod_{I=1}^3 ( 1+ \epsilon_I  w^{a_I h} )     \left[ 
{\rm tr}_{\rm D3} \gamma_{2h} 
-\frac{  w^{-{a_3 }h} {\rm tr}_{\rm D7} \gamma_{2h}  }{\prod_{I=1}^2 ( 1-w^{2 a_I h} )}  \right]  \int d\ell  \, { {\vartheta}_1\over  {\eta}^3} (i \ell+\ft12)    . \nn 
\eea 
Collecting the massless contributions from (\ref{tadpol}) one finds that 
$\tilde{\cal K}_h+\tilde{\cal A}_{2h}+\tilde{\cal M}_h$
form a complete square proportional to   
\bea
\left(   \prod_{I=1}^3 ( 1-  w^{2a_I h} )  {\rm tr}_{\rm D3} \gamma_{2h}    \,  + (w^{-2sh}-w^{2sh})   {\rm tr}_{\rm D7} \gamma_{2h}  
+ 4 \, \epsilon_0  \prod_{I=1}^3 ( 1+\epsilon_I  w^{a_I h} ) 
\right)^2 .
\label{tad0odd}
\eea
Using (\ref{chan}) and (\ref{id}) one can rewrite the combination  inside the brackets in 
(\ref{tad0odd}) as
\be
\tilde{\cal K}_h+\tilde{\cal A}_{2h}+\tilde{\cal M}_h \sim   \left[
\sum_{a=0}^{n-1} w^{2a h}  \mathcal I_a \right]^2, 
\label{tad1}
\ee
with
\be
{\mathcal I}_a= \sum_{b=0}^{n-1} ( I_{ab}\,N_b+J_{ab}\,M_b) +4  \epsilon_0 K_{a}\,.
\label{tadfinal0}
\ee
 We notice that ${\mathcal I}_a$ is precisely the anomaly associated to the gauge group U$(N_a)$ and is zero for $a=0,\ft{n}{2}$.
This shows that cancellation of local tadpoles ${\mathcal I}_a=0$ and of irreducible anomalies boil down to the same set of conditions
\be
I_{ab}\,N_b+J_{ab}\,M_b +4  \epsilon_0 \, K_{a} =0 \label{tadfinal}
\ee
for all $a$.  

\subsubsection*{Even $n$}

When $n$ is even,  we must distinguish between the tadpoles for fields $T_{2h}$ in the even twisted sectors, which propagate through the Klein, Moebius and Annulus amplitudes, and the odd ones $T_{2h+1}$ which only propagate along the Annulus.
Collecting all amplitudes contributing to the same tadpole one finds
\be
\tilde{\cal K}_{0,h}+\tilde{\cal K}_{0,h+n/2}+\tilde{\cal K}_{n/2,h}+\tilde{\cal K}_{n/2,h+n/2}+\tilde{\cal A}_{2h}+\tilde{\cal M}_h+\tilde{\cal M}_{h+n/2} = 0,
\qquad \tilde{\cal A}_{2h+1} = 0.
\ee
The two equations can be re-expressed as perfect squares:
\begin{align}
\label{tad0}%
&\left[\,\prod_{I=1}^3 ( 1-  w^{2a_I h} ) {\rm tr}_{\rm D3} \gamma_{2h} + (w^{-2sh}-w^{2sh}){\rm tr}_{\rm D7} \gamma_{2h}\right.+
\cr
&\qquad\qquad\qquad\qquad\qquad\:+\left.4\,\epsilon_0 \Big(\prod_{I=1}^3 (1+\epsilon_I w^{a_I h}) + \prod_{I=1}^3(1+\epsilon_I w^{a_I (h+n/2)})\Big)\right]^2
\cr
&\sim \left[\sum_{a=0}^{n/2-1} w^{2a h} ( {\mathcal I}_a + {\mathcal I}_{a+n/2})\right]^2=0,
\\[1.5em]
&\left[\,\prod_{I=1}^3 ( 1-  w^{a_I (2h+1)} ) {\rm tr}_{\rm D3} \gamma_{2h+1} + (w^{-s(2h+1)}-w^{s(2h+1)}){\rm tr}_{\rm D7} \gamma_{2h+1}\right]^2
\nn\\[.2em]
&\sim \left[\sum_{a=0}^{n/2-1} w^{a (2h+1)} ( {\mathcal I}_a - {\mathcal I}_{a+n/2})\right]^2 =0,
\end{align}
 where we used the fact that $K_a = K_{a+n/2}$.   Again cancellation of local tadpoles ${\mathcal I}_a=0$ matches the cancellation of anomalies in the quiver gauge theory.

For even $n$ another different orientifold projection can be achieved, called $\Omega_{\epsilon}'$ in the main text. This corresponds to the identification $\bar N_a = N_{\frac{n}{2}-a}$, $\bar M_a = M_{\frac{n}{2}-a}$. The cases with $n$ a multiple of four are in some sense ``trivial", since this $\hat{\Omega}_{\epsilon}$ coincides with the same orientifold projection previously described. The (\ref{omegagamma}) is now replaced by
\be\begin{aligned}
 {\rm tr}_{\rm D3} (\gamma^{-1}_{\Omega h}\gamma^{T}_{\Omega h})&=
     -\epsilon_0 (-1)^h\, {\rm tr}_{\rm D3} \gamma_{2h} =
     -\epsilon_0 w^{\frac{n}{2} h}\, {\rm tr}_{\rm D3} \gamma_{2h}\, ,\\
 {\rm tr}_{\rm D7} (\gamma^{-1}_{\Omega h}\gamma^{T}_{\Omega h})&=
    +\epsilon_0 (-1)^h\, {\rm tr}_{\rm D7} \gamma_{2h} =
     +\epsilon_0 w^{\frac{n}{2} h}\, {\rm tr}_{\rm D7} \gamma_{2h} \,,
\end{aligned}\ee
which can be easily checked for example in the $\mathbb{Z}_6$ case with $\gamma_{\Omega}$ explicitly given by (\ref{gampro2}).\\
 Such choice is allowed  because the extra phase squares to unity. This is a necessary condition to write again the sum of the transverse amplitudes as a perfect square, since the contributions to the Klein  and Annulus remain unchanged.
By performing the same steps as above, we recover the same identification of tadpole cancellation conditions with anomaly cancellation, as in (\ref{tad0}), but with a different expression for the orientifold contribution $K_a$:
\begin{equation}
\hat{K}_a = \sum_{I=1}^3  \epsilon_I (\delta_{2a,a_I+\frac{n}{2}} -\delta_{2a,\frac{n}{2}-a_I}  ).
\end{equation}
The example of this second projection on $\mathbb{C}^3/\mathbb{Z}_6$ was given in the main text.

\end{appendix}


\providecommand{\href}[2]{#2}\begingroup\raggedright\endgroup

\end{document}